 \documentclass[aps,pra,superscriptaddress,notitlepage,twocolumn,showpacs]{revtex4-1}
\usepackage[utf8]{inputenc}
\usepackage[usenames,dvipsnames]{xcolor}
\usepackage{amsmath}
\usepackage{amssymb}
\usepackage{mathtools}
\usepackage{esint}
\usepackage{dcolumn}
\usepackage{bm}
\usepackage{bbm}
\usepackage{blindtext}
\usepackage{braket}
\usepackage{natbib}
\usepackage{graphicx}

\usepackage[normalem]{ulem}
\usepackage{enumerate}
\usepackage{xpatch} 

\usepackage{multirow}
\usepackage[hidelinks=true,pdfborder={0 0 0},colorlinks=true,linkcolor=blue,urlcolor=blue,citecolor=blue]{hyperref}

\newcommand{\be}{\begin{eqnarray}}
\newcommand{\ee}{\end{eqnarray}}

\definecolor{darkgreen}{rgb}{0.0, 0.37, 0.0}

\definecolor{dgreen}{rgb}{0.0, 0.5, 0.0}

\renewcommand{\vec}[1]{\mathbf {#1}}
\newcommand{\veck}{\mathbf k}

\newcommand{\vecq}{\mathbf q}

\newcommand{\vecr}{\mathbf r}

\newcommand{\vecR}{\mathbf R}

\newcommand{\ced}{\hat c^\dagger}
\newcommand{\ce}{\hat c}

\begin{document}

\title{Theory of exciton-electron scattering in atomically thin semiconductors}

\author{Christian Fey}
\affiliation{Max-Planck-Institute of Quantum Optics, 85748 Garching, Germany}
\affiliation{Zentrum f\"ur Optische Quantentechnologien, Universit\"at Hamburg, Fachbereich Physik, 22761 Hamburg, Germany}
\author{Peter Schmelcher}
\affiliation{Zentrum f\"ur Optische Quantentechnologien, Universit\"at Hamburg, Fachbereich Physik, 22761 Hamburg, Germany}
\affiliation{The Hamburg Centre for Ultrafast Imaging, Universit\"at Hamburg, Luruper Chaussee 149, 22761 Hamburg, Germany}

\author{Atac Imamoglu}
\affiliation{Institute for Quantum Electronics, and Pauli Center for Theoretical Studies,   ETH Z\"urich, 8093 Z\"urich, Switzerland}

\author{Richard Schmidt}
\affiliation{Max-Planck-Institute of Quantum Optics, 85748 Garching, Germany}
\affiliation{Munich Center for Quantum Science and Technology (MCQST), Schellingstr. 4, 80799 M\"unchen, Germany}

\date{\today}
\begin{abstract}
The realization of mixtures of excitons and charge carriers in van-der-Waals materials presents a new frontier for the study of the many-body physics of strongly interacting Bose-Fermi mixtures. In order to derive an effective low-energy model for such systems, we develop an exact diagonalization approach based on a discrete variable representation that predicts the scattering and bound state properties of three charges in two-dimensional transition metal dichalcogenides. From the solution of the quantum mechanical three-body problem we thus obtain the bound state energies of excitons and trions within an effective mass model which are in excellent agreement with Quantum Monte Carlo predictions. The diagonalization approach also gives access to excited states of the three-body system. This allows us to predict the scattering phase shifts of electrons and excitons that serve as input for a low-energy theory of interacting mixtures of excitons and  charge carriers at finite density. To this end we derive an effective exciton-electron scattering potential that is directly applicable for Quantum Monte-Carlo or diagrammatic many-body techniques. As an example, we demonstrate the approach by studying the many-body physics of exciton Fermi polarons in transition-metal dichalcogenides, and we show that finite-range corrections  have a substantial impact on the optical absorption spectrum. Our approach can be  applied to  a plethora of many-body phenomena realizable in atomically thin semiconductors ranging from exciton localization  to induced superconductivity.

  \end{abstract}
\maketitle

\section{Introduction}
Interacting mixtures of fermions and bosons are at the heart of many paradigms of condensed matter physics, ranging from phonon and magnon-mediated superconductivity, mixtures of Helium-3 and Helium-4, polaron mobility, to electrons coupled to dynamical gauge fields. Recent progress in the trapping and manipulation of ultracold quantum gases made cold atoms a promising platform to study physics of strongly interacting quantum mixtures \cite{Bloch2008,Lous2018}. As a key aspect these systems feature bosons that do not  appear as collective excitations of the many-body system, such as magnons or phonons, but instead represent point-like particles which interact  with the fermions by coupling terms that are non-linear in their creation operators. Exploiting this fact made it possible to realize interactions of bosons and fermions in the strong-coupling regime that goes beyond the paradigm of the Fr\"ohlich  model \cite{Froh3,Froh4,Rath2013}, leading to the recent observation of strong coupling Bose polarons \cite{Hu2016,Jorgensen2016,Camargo2018}.

In contrast, typical solid state realizations of Bose-Fermi mixtures, concern pointlike fermions (electrons) that interact with bosonic degrees of freedom which are collective, low-energy excitations of either the crystal lattice (phonons) or the electronic system itself (e.g. plasmons or magnons). In order to realize a good representation of \textit{point-like} bosons one faces the challenge to ensure that the density of the fermions, as characterized by their Fermi energy $\epsilon_F$, remains sufficiently dilute as compared to the extent of the bosonic particle which is characterized, for instance, by its binding energy or internal excitation energies. While this condition is well-satisfied in cold atoms, where typical Fermi energies are on the order of $\sim$h~$\times$~kHz, and thus tiny compared to atomic transition frequencies, $\sim$THz, the creation of such a large scale separation is a key challenge for the solid-state realization of atom-like Bose-Fermi mixtures. 

One prime example for atom-like bosons in solid state matter are excitons, which allowed for the realization of Bose-Einstein condensation of excitons and the observation of superfluidity \cite{Imamoglu1996,Kasprzak2006,Carusotto2013}. In  order to promote these systems to Bose-Fermi mixtures the semiconductor can be doped with charge carriers. Excitons in \textit{bulk} semiconductors are bound by a binding energy on the order of 10 meV \cite{Deng2010}. Fermi energies of interest are, however, of the same order which invalidates the picture of well-defined Bose-Fermi mixtures.

Atomically thin transition metal dichalcogenides offer a way around this limitation. Indeed, in the last two decades the ingeniously simple process of mechanical exfoliation allowed to explore the vast playground of van der Waals materials ranging from gapless graphene \cite{Novoselov2004,Zhang2005,Geim2007}, large band gap insulators \cite{Novoselov2005}, superconductors \cite{Xi2015}, twisted bilayer graphene \cite{Bistritzer2011,Cao2018,Cao2018b,Yankowitz2019}, and ferromagnets \cite{Huang2017,Bonilla2018}. With transition metal dichalcogenides a new class of atomically thin semiconductors with potential technological applications has emerged \cite{Mak2012}  that provides a novel platform to realize strongly interacting mixtures of point-like bosons and fermions. In contrast to their bulk counterparts, in atomically thin materials screening of Coulomb forces is reduced owing to the absence of an all-encompassing dielectric environment. This leads to the existence of tightly bound excitons with a binding energy $\epsilon_X$ on the order of hundreds of meV \cite{chernikov2014}. As a consequence it is possible to reach the desired regime of large energy separation  where excitons  remain well-defined atom-like particles even in the presence of a substantial electron Fermi energy, i.e. $\epsilon_F/\epsilon_X\ll1$. 
Moreover, the existence of a trion bound state with binding energy $\epsilon_T\approx 30$~meV opens a window to the  strong coupling regime, where the interaction energy, characterized by $\epsilon_T$, competes with the kinetic energy of the charge-carrier Fermi gas, i.e. $\epsilon_F/\epsilon_T\approx 1$. Besides the potential technological applications ranging from light emitting diodes \cite{Britnell2013} to solar cells \cite{Furchi2014}, these features make  TMDs a serious new competitor to cold atomic systems as a platform to study paradigm many-body models of condensed matter theory in a controlled, nanoscopic environment.

First examples that explored the rich physics of strongly interacting Bose-Fermi mixtures of excitons and electrons in 2D semiconductors addressed the regime of low boson density \cite{Sidler2016,Ravets2018}. Here the physics of Fermi polarons \cite{Suris2003,Prokofiev_FrohPolaron1,Prokofiev_FrohPolaron2,Punk2009}, single \textit{mobile} quantum impurities immersed in a Fermi gas, is realized \cite{Schmidt2012b,Efimkin_Many-body_2017,Efimkin_Exciton-polarons_2018,cotlet2018,schmidt2017} which has been a long-standing problem in theoretical physics that touches upon questions about the existence of quasiparticles \cite{Anderson1967,AchimKopp} and fundamentals of transport \cite{Kondo1983}. More recently, it was demonstrated that the scattering of electrons and excitons provides a new pathway towards cooling of exciton-polaritons leading to enhanced optical gain of 2D materials \cite{Tan2019}. Exploiting further the interactions of electrons mediated by exciton exchange has been proposed to enable induced superconductivity \cite{Cotlet2016,Kavokin2016} and the realization of supersolids  \cite{Shelykh2010}.

In order to obtain a  reliable theoretical description of the physics of Bose-Fermi mixtures in TMDs, an effective model of the scattering physics of electrons and excitons is paramount. On the one hand, such a low-energy description should be sufficiently simple to be a viable input for many-body techniques ranging from quantum Monte Carlo to diagramatics. On the other hand, the interaction model has to provide a quantitatively accurate description. Since the relevant many-body scales --- set by the Fermi energy, exciton density or temperature --- are substantially smaller than the exciton energy one would like to derive a model where high-energy scales down to the exciton energy have been integrated out, so that only a direct interaction between excitons and electrons has to be considered.

In this work we use exact diagonalization to derive an effective, accurate interaction model for excitons and electrons in transition metal dichalcogenides. To this end we solve exactly the quantum mechanical problem of three charge carriers in TMDs in an effective mass model. Using a discrete variable representation and exploiting the tensorial structure of the kinetic part of the three-body Hamiltonian our approach yields trion energies that are in excellent agreement with QMC calculations. Moreover, we find exotic excited trion bound states, not previously discussed in the literature, and which correspond to the binding of electrons to Rydberg excitons in a p-wave configuration where the constituent particles possess opposite angular momenta. 

Most importantly, however,  our approach also gives access to the structure of three-body envelope wave functions as well as scattering states above the trion dissociation threshold. From this we show that the picture of exciton-electron scattering and thus the description in terms of Bose-Fermi mixtures is well-justified. The scattering physics of excitons and electrons is universally captured by the energy dependent 2D scattering phase shift $\delta(E)$. We extract $\delta(E)$ directly from the full spatial structure of the three-body wave functions by including up to $10^6$ basis states in the exact diagonalization. The  results  show that contact interaction models for excitons and electrons are insufficient for many key observables such as polaron energies or transition temperatures to superconducting phases  induced by exciton exchange.

\begin{figure}[t]
\includegraphics[width= 1\linewidth]{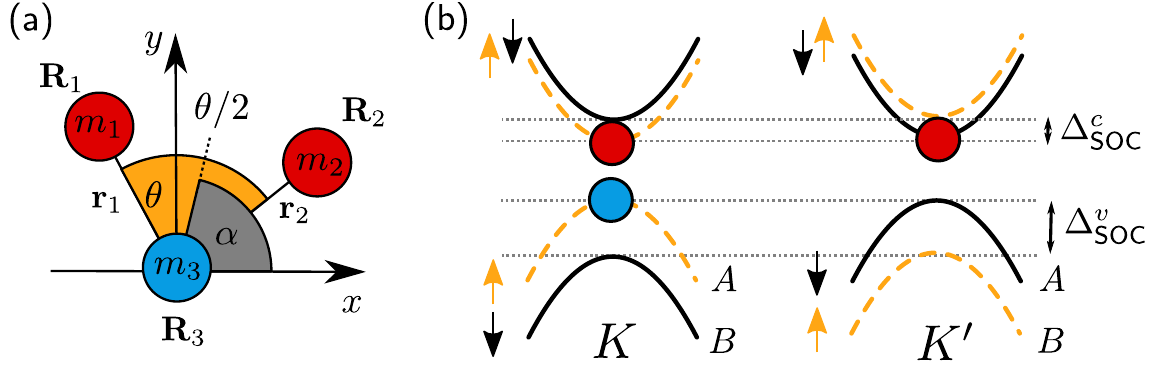}
\caption{(a) Parametrization of the three-body system with radial degrees of freedom $r_1=|\vecr_1|$, $r_2=|\vecr_2|$ and the relative angle $\theta$. The  angle $\alpha$ is defined with respect to the angle bisector of $\theta$. It describes the orientation of the trion in the $xy$-plane and corresponds to the total angular momentum of the three-body complex. The masses of the charge carriers are $m_1$, $m_2$ and $m_3$. (b) Schematic illustration of an intervalley trion in the band structure around the $K$- and $K'$-points. The spin-orbit splitting of the valence bands $\Delta^v_\text{SOC}$ is significantly larger than the splitting of the conductions bands $\Delta^c_\text{SOC}$.}
\label{fig:coordinates}
\end{figure}

The work is structured as follows. In Section \ref{SecModel} we introduce the Hamiltonian that describes the motion and interactions of three charge carriers in two-dimensional TMD in an effective mass approximation, and we detail how the exact diagonalization approach is applied. In Section \ref{Sec_Ground&Excited} we focus on the analysis of the trion ground state and the structure of its envelope wave function. We then discuss the excitation spectrum of the system including excited trion states and scattering states. Section \ref{Sec_ElectronXScatt} is devoted to the calculation of the scattering phase shifts of electrons and excitons and the derivation of an effective low-energy model for exciton-electron interactions. In Section~\ref{Sec_ElectronDefectScatt} we demonstrate the applicability of this interaction model by analyzing the optical absorption spectra of n-doped MoSe$_2$. We summarize our findings in Section~\ref{SecSummary} and outline future directions.

\section{Effective mass model}\label{SecModel}

To describe exciton-electron scattering and the properties of trions in atomically thin semiconductors, we employ an effective mass model for three charged point-like particles in two dimensions.  Each particle has a coordinate $\vecR_i$, a parabolic band mass $m_i$ and carries a charge $q_i$ ($i=1,2,3$); for an illustration see Fig.~\ref{fig:coordinates}(a).
Similar to the studies \cite{berkelbach_theory_2013, Courtade_Chraged_excitons_2017} we introduce relative coordinates $\vec{r}_1$ and $\vec{r}_2$ that describe the spatial relative vectors between the particles $i=1,2$ and the particle $i=3$. The center-of-mass motion can be separated and the remaining Hamiltonian for the internal three-body dynamics reads

\begin{equation}
\begin{split}
\hat{H} &= -\frac{1}{2\mu_1} \Delta_{\vec{r}_1}  -\frac{1}{2\mu_2} \Delta_{\vec{r}_2} -\frac{1}{2 m_3} \vec{\nabla}_{\vec{r}_1} \cdot \vec{\nabla}_{\vec{r}_2}  \\
&+  q_1q_3 V_\text{K}(r_1) +  q_2q_3 V_\text{K}(r_2) + q_1 q_2 V_\text{K}(|\vec{r}_1-\vec{r}_2|) ,
 \end{split}
\label{eqn:H_valence}
\end{equation}
where $\mu_i= m_i m_3/(m_i+m_3)$ are the reduced masses.

The interactions among the charge carriers are modeled with the Keldysh potential \cite{keldysh1979, Cudazzo_dielectric_screening_2011, berkelbach_theory_2013} 
\begin{equation}
V_\text{K}(r)= \frac{\pi}{r_0} \left[H_0(r / r_0)- Y_0(r /r_0)\right],
\label{eqn:Keldysh potential}
\end{equation}
where $H_0$ and $Y_0$ are the  Struve function and the Bessel function of the second kind, and  the screening length $r_0=2\pi \chi_{2D}$ is linked to the 2D polarizability $\chi_{2D}$ of the planar material. Eq.~\eqref{eqn:Keldysh potential} describes to a good approximation deviations from a Coulomb potential at short distances arising due to dielectric screening, while at large distances the Coulomb's law is recovered, $V(r) \to 1/r$.  Note, while further corrections to Eq.~\eqref{eqn:Keldysh potential} exist, we restrict us here to this specific form in order to enable a direct comparison of our results for trion and exciton energies with QMC calculations \cite{kylanpaa_binding_2015}.

For the effective masses and the screening lengths we employ material parameters obtained from DFT band structure calculations \cite{kylanpaa_binding_2015} as stated in Tab.~\ref{tab:table1}. A corresponding sketch of the band structure in TMDs around the energetically degenerate $K$ and $K'$ points of the Brillouin zone is presented in Fig.~\ref{fig:coordinates}(b). The two valence bands A and B are subject to a spin-orbit splitting of approximately 100 meV and the effective band masses of charge carriers in these bands are significantly different. The splitting of the conduction band  $\Delta_\text{SOC}^c$ is roughly 10-100 times smaller than  $\Delta_\text{SOC}^v$ and also the mass difference is much less pronounced so that we use an electron mass $m_e$ that is averaged over these two, almost degenerate bands. As illustrated in Fig.~\ref{fig:coordinates}(b) we focus on configurations that are composed of two electrons and a single hole. 

\begin{table}[t]
\begin{tabular}{|l|l|l|l|l|}\hline  & MoS$_2$ & MoSe$_2$& WS$_2$&WSe$_2$ \\ \hline 
{\bf material parameter\cite{kylanpaa_binding_2015} }& & & &\\
$r_0$ (\AA) &44.6814 &53.1624 & 40.1747& 47.5701 \\
$m_e$ &0.47& 0.55 & 0.32 & 0.34\\
$m_h$, A-band &0.54& 0.59 & 0.35 & 0.36\\ \hline
{\bf exciton energy}& & & &\\
present work (meV) & 526.0&  476.7 & 508.6 & 456.0 \\
QMC \cite{kylanpaa_binding_2015} (meV) & 526.5(2)& 476.9(2) &509.8(2)& 456.4(2) \\ \hline
{\bf mobile trion energy}& & & &\\
present work (meV)& 31.7&  27.7  & 34.2 & 28.4 \\
QMC \cite{kylanpaa_binding_2015} (meV)& 32.0(3)& 27.7(3) & 33.1(3) & 28.5(3) \\ \hline
\end{tabular}
\caption{Exciton and trion binding energies for various TMDs obtained from exact diagonalization. Material parameters are taken from DFT computations \cite{kylanpaa_binding_2015}. Energies are compared to the path-integral Monte-Carlo simulations  presented in Ref.~\cite{kylanpaa_binding_2015}.}
\label{tab:table1}
\end{table}

The Hamiltonian (\ref{eqn:H_valence}) can be further simplified by introducing the polar coordinates $r_1=|\vecr_1|$, $r_2=|\vecr_2|$, $\theta$ and $\alpha$ which parametrize the coordinates $\vec{r}_1$ and $\vec{r}_2$, see Fig. \ref{fig:coordinates}.
In these coordinates we express the wave function in the form
\begin{equation}
\psi(r_1,r_2,\theta,\alpha)= \frac{u(r_1,r_2,\theta)}{\sqrt{2 \pi r_1 r_2}} \exp(i m \alpha)
\end{equation}
where  $u(r_1,r_2,\theta)$ is normalized as
\begin{equation}
\int \limits_{r_1=0}^\infty \int \limits_{r_2=0}^\infty \int \limits_{\theta=0}^{2 \pi}  dr_1 dr_2 \ d\theta |u(r_1,r_2,\theta)|^2 = 1 .
\end{equation}
Since Eq.~\eqref{eqn:H_valence} is invariant under in-plane rotations described by the angle $\alpha$, the angular momentum $m$ is conserved. In this work we focus exclusively on $m=0$, for the resulting, reduced Hamiltonian see App.~\ref{sec:appendix}.

To compute the eigenstates of the Hamiltonian \eqref{eqn:H_valence} we follow an exact diagonalization scheme. To this end we construct the Hamiltonian  in a discrete variable representation (DVR) for each degree of freedom (DOF). As basis functions for the radial DOF, e.g.~$r_1$ (analogous for $r_2$), we employ $\phi_n(r_1)=\sqrt{r_1/(l_0 n)} \exp(-r_1/(2l_0))L^1_{n-1}(r_1/l_0)$ with the generalized Laguerre Polynomials $L^1_{n-1}(r_1)$, $n \in \mathbb{N}$, and a length scale parameter $l_0$ that controls the spatial resolution \cite{mccurdy_resonant_2003}. The radial basis functions satisfy the boundary condition $\phi_n(r_1) \propto \sqrt{r_1} $ for $r_1\to 0$.  For the angular variable $\theta$ the basis functions $\sqrt{1/2 \pi} \exp(i l \theta)$ with $l \in \mathbb{Z}$ satisfy $2\pi$ periodic boundary conditions.

Starting from these basis functions, we follow the DVR approach  (for a review see \cite{beck_multiconfiguration_2000}), and  diagonalize the position operators $\hat r_1$, $\hat r_2$ and $\cos(\hat \theta/2-\theta_0)$ with an appropriately chosen offset $\theta_0$. This procedure leads to a new set of wave packet basis states that are strongly localized on a spatial grid and thus provide a discrete representation of position space.  The potential $\hat{V}$ is diagonal in this new basis and can therefore be evaluated efficiently. Note that while the angular grid is spaced equidistantly, the radial grid becomes increasingly dense at short distances which is beneficial to resolve the short-range structure of the three-body complexes in great detail. The extents of the radial grids $r_1^\text{max}$ and $r_2^\text{max}$ are determined by $l_0$ as well as the size of the radial basis set. For each DOF, we typically employ 60 basis functions; for checks of convergence, however, a total of up to $10^6$ basis states is included.

The obtained eigenfunctions $u(r_1,r_2,\theta)=u(\vec{r}_1,\vec{r}_2)$ can be interpreted as the envelope functions of the Bloch solution of the three-body system in the crystal $\psi_{S_1,S_2,S_3}(\vec{R}_1,\vec{R}_2,\vec{R}_3)$ where
the collective index $S_i$ characterizes the charge carriers in the band structure. For instance, the negatively charged intervalley trion depicted in Fig.~\ref{fig:coordinates}(b) has $S_1=\{K,\uparrow\}$, $S_2=\{K',\downarrow\}$ and $S_3= \{K,\uparrow\}$.
Taking into account spin statistics, its Bloch state can be approximated as  \cite{Courtade_Chraged_excitons_2017}
\begin{equation}
\begin{split}
&\psi_{S_1,S_2,S_3}(\vec{R}_1,\vec{R}_2,\vec{R}_3)= \frac{ e^{i\vec{K}_0 \vec{R}_0}}{\mathcal{N}} \mathcal{U}_{S_3}(\vec{R}_3) \times \\
&\left[u(\vec{r}_1,\vec{r}_2) \mathcal{U}_{S_1}(\vec{R}_1)\mathcal{U}_{S_2}(\vec{R}_2)- u(\vec{r}_2,\vec{r}_1)\mathcal{U}_{S_2}(\vec{R}_1)\mathcal{U}_{S_1}(\vec{R}_2) \right] 
\end{split}
\label{eqn:Trion_wave_function}
\end{equation} 
with the normalization constant $\mathcal{N}$, the single particle Bloch functions $\mathcal{U}_{S_i}(\vec{R}_i)$, and the center-of-mass coordinate and wave vector  $\vec{R}_0$ and $\vec{K}_0$, respectively. Although not stated explicitly in Eq.~\eqref{eqn:Trion_wave_function}, it is implied that the envelop $u(\vec{r}_1,\vec{r}_2)$ depends also on the combined spin and valley indices $S_i$, i.e. for a given set $S_i$, one determines $u(r_1,r_2,\theta)$ based on Eq.~\eqref{eqn:H_valence} with corresponding effective masses $m_i$. Note, in the present work, in order to make direct comparison to state-of-the-art QMC predictions \cite{kylanpaa_binding_2015}, we do not take into account short-range Coulomb-exchange \cite{Glazov_2015, Plechinger_2016,Courtade_Chraged_excitons_2017} as well as the non-zero Berry curvature in TMD structures \cite{srivastava2015}. However, both effects can be included in our approach. 


\section{Excitons, trions and their excitation spectrum}\label{Sec_Ground&Excited}
First we study excitons and trions which are the  ground states of the two- and three-body problem, respectively. Specifically, we focus on the example of negatively charged trions that consist of one hole and two electrons. Depending on the spin and valley index, the hole is situated in either the A or B valence band, leading to $A$ and $B$ trions (and excitons). For simplicity we focus here exclusively on holes in the energetically higher $A$-band; for an illustration see Fig.~\ref{fig:coordinates} (b).
Since for equal conduction band masses the Hamiltonian (\ref{eqn:H_valence}) is invariant under exchange of $\vec{r}_1$ and $\vec{r}_2$, one can choose a basis of eigenstates $u(r_1,r_2,\theta)$ that are either symmetric $u(r_1,r_2,\theta)=u(r_2,r_1,-\theta)$ or antisymmetric functions $u(r_1,r_2,\theta)=-u(r_2,r_1,-\theta)$.
This symmetry is closely related to the electron spin degrees of freedom present in the total wave function in Eq.~\eqref{eqn:Trion_wave_function}, which, owing to spin statistics, is, by construction, antisymmetric under electron exchange.  
For the particular configuration presented in Fig.~\ref{fig:coordinates}(b), symmetric (antisymmetric) wave functions $u(r_1,r_2,\theta)$ thus correspond to electron-spin singlet (triplet) states.

\textbf{Ground states.---} In our simulations the trion ground state is always spatially symmetric and we do not find zero-angular-momentum states with an antisymmetric envelop $u(r_1,r_2,\theta)$ below the exciton line. This agrees with results based on variational wave functions that predict spatially antisymmetric trions only for non-zero angular momentum \cite{Courtade_Chraged_excitons_2017,sergeev2001triplet}. The resulting binding energies of  excitons and trions are presented in Table \ref{tab:table1} for different classes of TMD. All energies are in excellent agreement with path-integral Monte Carlo simulations \cite{kylanpaa_binding_2015}, and the predicted trion energies lie within the range of experimental results \cite{wang2018}. An even more accurate agreement with experiments can be achieved by, for instance, accurately incorporating the influence of the dielectric environments in TMD heterostructures \cite{chernikov2014,robert2018,florian2018}.

\begin{figure}[t]
\includegraphics[width= 0.99 \linewidth]{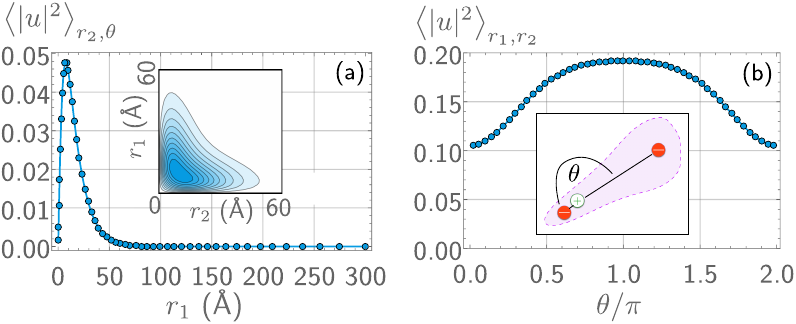}
\caption{Charge-carrier density of the MoS$_2$ ground state trion. (a) The reduced density $\left<|u|^2\right>_{r_2,\theta} (r_1)$ obtained from the average over variables $r_2$ and $\theta$ (blue circles). The inset depicts the probability density of the radial configuration $\left<|u|^2\right>_{\theta}(r_1,r_2)$. (b) The probability density of the angular configuration $\left<|u|^2\right>_{r_1,r_2}(\theta)$ with an illustration of the spatial structure of the trion as inset.
}
\label{fig:mobile_ground}
\end{figure}

We now turn to the study of the real-space structure of the three-body wave function. In models describing the many-body physics of excitons and electrons the trion is typically  regarded as the bound state of an exciton and an additional charge carrier \cite{Sidler2016,Efimkin_Many-body_2017,Efimkin_Exciton-polarons_2018}. This picture can be tested with our approach where the full spatial structure of the three-body wave function $u(r_1,r_2,\theta)$ is accessible. In order to visualize the dependence of this wave function  on its three variables, we show reduced densities that are obtained by averaging $|u(r_1,r_2,\theta)|^2$ over either radial or angular coordinates. As an example, we show the predicted trion ground state  density for MoS$_2$ in Fig. \ref{fig:mobile_ground}.

In the formation of the trion the two electrons compete for the tight binding with the hole. This becomes evident in the inset of Fig.~\ref{fig:mobile_ground}~(a)  showing the reduced charge-carrier density $\left<u\right>_{\theta} (r_1,r_2)$ after an average over the angular coordinate $\theta$. Although electron-hole separations are most likely around $r_1=r_2 \sim 15$~{\AA}, it is also possible to have large separations in one coordinate, e.g. $r_1\sim 40$~{\AA}, under the condition of tight binding in the other coordinate, e.g. $r_2\sim 10$~{\AA}. 

Performing an additional average over the coordinate $r_2$ one arrives at the probability density $\left<u\right>_{r_2,\theta} (r_1)$ of having one electron-hole pair at a  separation $r_1$. As shown in  the main panel of Fig.~\ref{fig:mobile_ground}(a), this density is peaked at around 10~{\AA}. Its first moment $\left<r_1\right> \approx 16$~{\AA} provides an estimate for the spatial extent of the trion, which is larger than the mean binding length of the corresponding $X_{1s}$ exciton $\sim$ 10~\AA.

In addition to these radial properties, the angular structure of the trion is characterized by the angular density $\left<u\right>_{r_1,r_2}(\theta)$ presented in Fig.~\ref{fig:mobile_ground}~(b). On top of an isotropic background distribution with $1/(2 \pi) \approx 0.16$ the density is peaked around the linear configuration with $\theta =\pi$ and suppressed around $\theta=0$. This is a consequence of  Coulomb interactions that leads to a polarization of the tightly bound excitonic substructure of the trion due to the presence of the additional charge carrier (see illustration in Fig.~\ref{fig:mobile_ground}). \\

\begin{figure}[t]
\includegraphics[width= \linewidth]{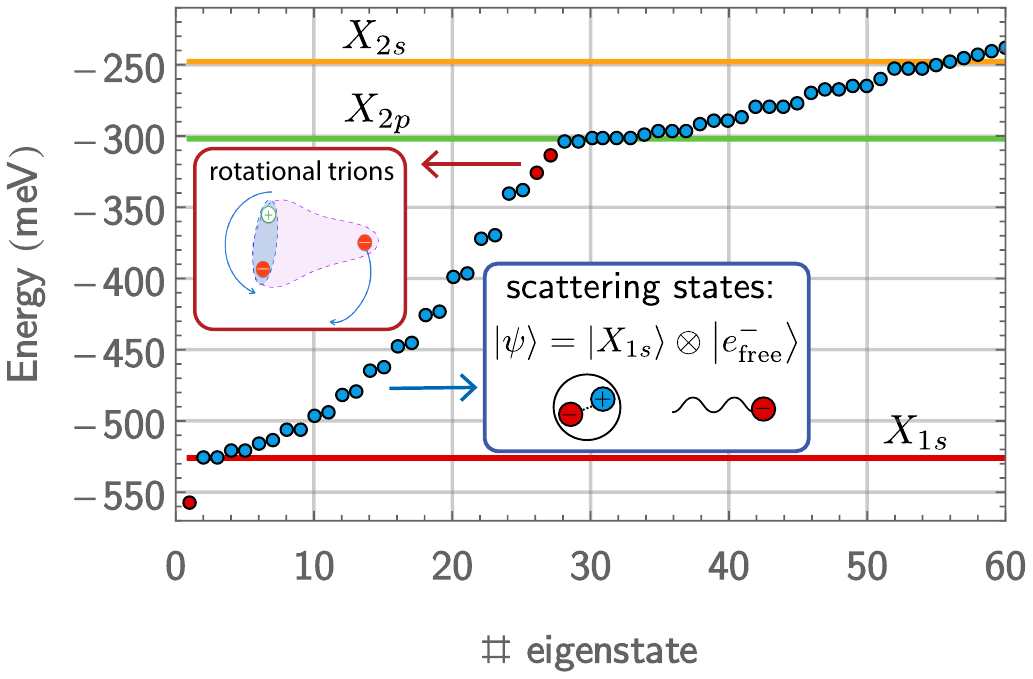}
\caption{Energy spectrum of the lowest 60 eigenstates for the MoS$_2$ three-body system obtained from exact diagonalization on a grid of length $r_1^\text{max}=r_2^\text{max}= 300 \text{ \AA}$. Red (blue) dots correspond to bound trions (exciton-electron scattering states, see pictorial representation in the inset). The colored horizontal lines show the energies of the  MoS$_2$ excitons including Rydberg states of low angular momentum. 
}
\label{fig:mobile_spectrum}
\end{figure}

\textbf{Excited states.---} Having discussed the  trion ground state of the three-body system, we turn next to excited states. Fig.~\ref{fig:mobile_spectrum} presents the energy spectrum of the  lowest 60 eigenstates of the three-body Hamiltonian for  MoS$_2$. The colored horizontal lines indicate the energy of the $X_{1s}$ exciton as well as the two energetically lowest exciton Rydberg states $X_{2p}$ and $X_{2s}$. The binding energy of the trion $\sim$ 32 meV appears here as the energy difference between the lowest three-body state and the 1$s$ exciton.

\begin{figure}[t]
\includegraphics[width= 0.99 \linewidth]{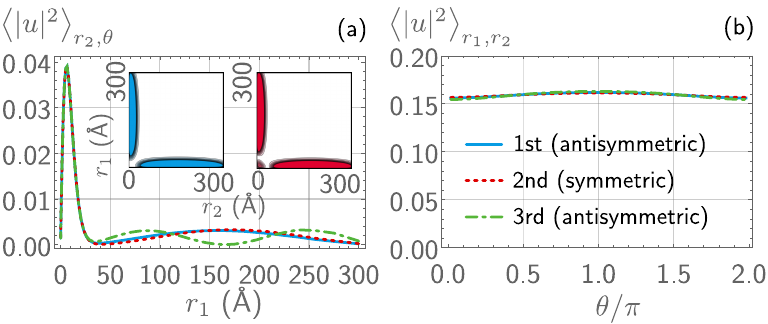}
\caption{Probability densities of the three energetically lowest exciton-electron scattering states of the MoS$_2$ three-body system. (a) Reduced probability distribution $\left<|u|^2\right>_{r_2,\theta}$ as function of the electron-hole separation $r_1$. The 3rd state (dashed dotted green line) displays one more radial node than the 1st and 2nd state (solid blue and dashed red line, respectively). The latter can be distinguished via their symmetry under electron exchange. This becomes evident in the inset contour plots of the radial correlation $\left<|u|^2\right>_{\theta}(r_1,r_2)$ (1st state (left) and 2nd state (right)). Due to its symmetry the antisymmetric 1st state has a vanishing density along the diagonal. (b) The angular distributions $\left<|u|^2\right>_{r_1,r_2}(\theta)$ of the three states are almost identical and nearly homogeneous (on this scale).}
\label{fig:mobile_2nd}
\end{figure}

The energies of the next higher eigenstates lie above the $X_{1s}$ exciton energy. As shown  in Fig.~\ref{fig:mobile_2nd}(a) for the 2nd and 4th eigenstate, these states are not bound, which is reflected in the fact that their radial densities $\left<u\right>_{r_2, \theta}(r_1)$ do not decay exponentially with $r_1$ independently of  system size. The nature of these excited states becomes evident from the fact that electrons are only weakly correlated: as can be seen from the inset in Fig. \ref{fig:mobile_2nd}(a), one electron is  close to the hole, while the other electron is very distant and delocalized.  As illustrated in the inset in Fig.~\ref{fig:mobile_spectrum}, these states thus correspond  to scattering states of a quasi-free electron which scatters off the 1$s$ exciton. The latter statement is further supported by the fact that the electron density at short distances closely resembles the density profile of a single 1s exciton state in absence of an additional charge carrier. Moreover, the angular densities in Fig.~\ref{fig:mobile_2nd}(b) are nearly homogeneous and exhibit only small polarization effects. Due to the finite extent of the radial grids (here 300 {\AA}) the energies of the scattering states are discrete. However, in the limit of very large spatial grids  the spectrum becomes dense and one recovers the quadratic dispersion relation of the scattered electron (as already visible in Fig.~\ref{fig:mobile_spectrum}).
In addition, the scattering states appear always as doublets of energetically almost degenerate states. Each doublet has the same number of radial nodes but different spatial symmetry under coordinate exchange (antisymmetric vs. symmetric). Their symmetry can be readily obtained by evaluation of the character of the numerical wave functions $u(r_1,r_2,\theta)$. For instance, the 2nd eigenstate in Fig.~\ref{fig:mobile_2nd}(a) has one radial node and is symmetric. Its radial density (inset with red filling) is finite close to $r_1=r_2=0$ which is necessarily forbidden for antisymmetric states as can be seen for the 1st excited, antisymmetric  eigenstate (inset with blue filling).

\begin{figure}[t]
\includegraphics[width= 0.99 \linewidth]{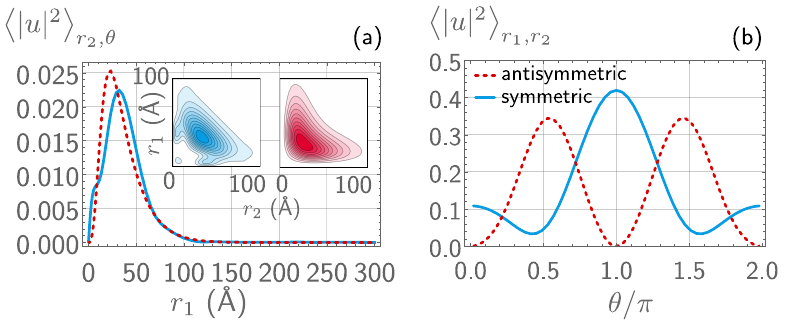}
\caption{Rotational MoS$_2$ trion with spatially symmetric and antisymmetric wave functions (solid blue vs. dashed red lines). The reduced probability densities are obtained by averaging out one or two degrees of freedom. (a) Probability density $\left<|u|^2\right>_{r_2,\theta} (r_1)$ as function of electron-hole separation $r_1$. The insets displays the radial correlations $\left<|u|^2\right>_{\theta}(r_1,r_2)$ for the spatially symmetric (left) and antisymmetric (right) state. (b)  Corresponding angular probability density $\left<|u|^2\right>_{r_1,r_2}(\theta)$.}
\label{fig:mobile_26th}
\end{figure}

As the eigenenergies approach the energy of the $X_{2p}$ Rydberg exciton at $E\approx 300$ meV, the quadratic dispersion relation becomes modified. Here an additional scattering channel  opens up that corresponds to the scattering between an $X_{2p}$ exciton  and an electron of finite angular momentum. It turns out that these scattering states represent the dissociation continuum of two new trion bound states that appear in the spectrum.
These bound states, shown as red dots in Fig.~\ref{fig:mobile_spectrum}, lie approximately 25 meV (11 meV) below the $X_{2p}$ exciton and have a symmetric (antisymmetric) wave function $u(r_1,r_2,\theta)$. Their bound state character is visible in the exponential envelop in the reduced densities  shown in Fig.~\ref{fig:mobile_26th}(a).
Moreover, as can be seen from the angular densities in Fig.~\ref{fig:mobile_26th}(b), these states are excited along the $\theta$ direction. Since the \textit{total} angular momentum $m$ is zero, these states can be regarded as $2p$ trions composed of a rotating electron that is bound to a counter-rotating $X_{2p}$ exciton. 

A special property of the antisymmetric $2p$ trion is that it is the energetically lowest state satisfying $u(r_1,r_2,\theta)=-u(r_1,r_2,-\theta)$. Parity with respect to $\theta$ is a subsymmetry of the Hamiltonian and the antisymmetric $2p$ trion is, consequently, the ground state of the odd parity sector. For this reason it is protected against couplings to continuum states and corresponding decay processes. In contrast, the symmetric $2p$ trion has even parity under $\theta$ reflections and possesses a finite admixture of the $X_{1s}$ state. This admixture is visible as a small enhancement of the radial density at short distances and it contributes to the characteristic shape of the density $\left<|u|^2\right>_\theta(r_1,r_2)$ resembling a devil-fish silhouette. As a consequence of the resulting coupling to continuum states, it is expected that the lifetime of the symmetric $2p$ trion will be decreased.    

Similarly to the excited trions, the exciton-electron scattering states above the $X_{2p}$ threshold have a mixed excitonic $X_{1s}$, $X_{2p}$ and even $X_{2s}$ character. This indicates that non-elastic scattering processes between the corresponding asymptotic scattering states are possible, similar to the collisions of rovibrationally excited molecules. This highlights TMDs as a new frontier to emulate the physics of molecular collisions in two dimensions in a solid-state setting.  We note that in our calculations we do not find evidence for stable $2s$ trions, i.e. bound state between a $X_{2s}$ exciton and an electron of zero angular momentum \cite{Arora_2019}. This is consistent with previous studies showing that negatively charged $2s$ trions are only stable if $m_e>m_h$ when considering pure Coulomb interactions in 2D \cite{Shiau_Combescot_Chang_2012}.

\section{Electron - exciton scattering}\label{Sec_ElectronXScatt}

\begin{figure*}[t]
\includegraphics[width= 0.99 \linewidth]{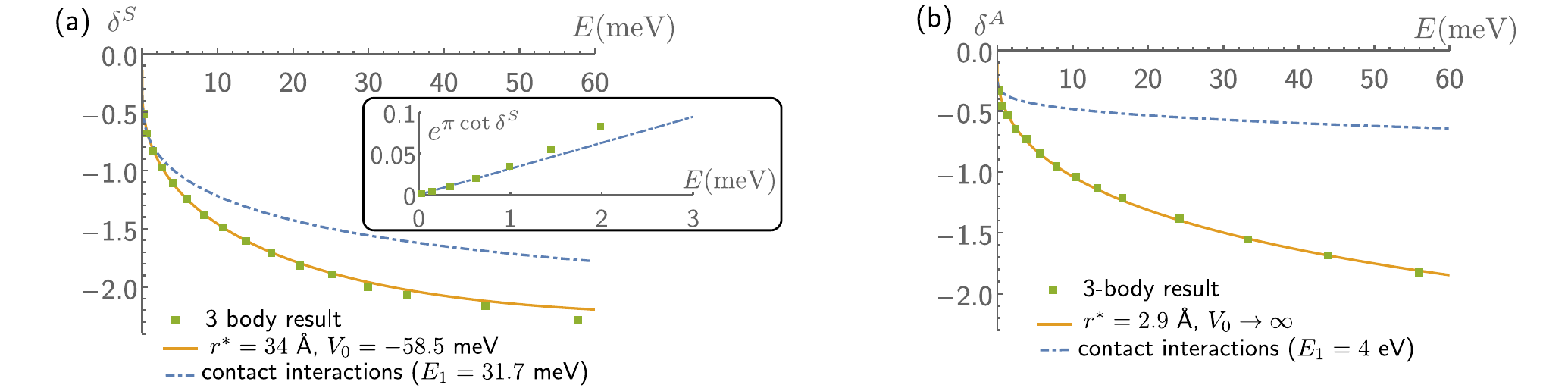}
\caption{Energy-dependent symmetric phase shifts $\delta^S (E)$ (a) and antisymmetric phase shifts $\delta^T (E)$ (b) for exciton-electron scattering in MoS$_2$. Numerical three-body results  from exact diagonalization (green squares) are compared to the phase shifts resulting from the effective exciton-electron pseudo-potential Eq.~\eqref{eqn:pseudopotential} (solid orange line) with parameters $r^*$ and $V_0$ given in Table \ref{tab:table2}.
Additionally, phase shifts for contact interactions, see Eq.~\eqref{eqn:delta_lowe}, are shown as dashed-dotted blue lines with $E_1=31.7$~meV (a) and $E_1=4$~eV (b). These parameters have been chosen to match the low-energy scaling of the three-body results. This is illustrated in the  inset in (a) that shows the rescaled phase shifts $\exp{(\pi \cot{\delta^{S}})}$ which are expected to scale $\propto E$ in the limit of very small energies, see Eq.~\eqref{eqn:delta_lowe}.  Only for the symmetric channel $E_1$ is related to the binding energy of the trion, whereas, for the symmetric channel, it does not correspond to any bound-state property.}
\label{fig:phaseshifts_mobile}
\end{figure*}

As discussed in Sec.~\ref{Sec_Ground&Excited} the energetically low-lying excited states  of the three-body Hamiltonian above the $X_\text{1s}$ exciton line correspond to electrons with zero angular momentum that are scattered off $X_\text{1s}$ excitons. Compared to free electrons their wave functions are subject to an s-wave scattering phase shift induced by an effective exciton-electron interaction. In consequence, the radial densities $\left<|u|^2\right>_{r_2,\theta} (r_1)$ of these s-wave scattering states with an energy $E>0$ relative to the exciton energy satisfy for large $r_1$ 
\begin{equation}
\left<|u|^2\right>_{r_2,\theta} (r_1) \approx k r_1 \left[\alpha(k) J_0(k r_1) + \beta(k) Y_0(k r_1)  \right]^2.
\label{eqn:free_particle_2d}
\end{equation} 
Here $J_0$ and $Y_0$ are the Bessel functions of the first and second kind and the wave number $k$ satisfies  $E= k^2/(2M_\text{red}) $ where $M_\text{red}= m_e m_X/(m_e+m_X)$  is the reduced mass of the exciton-electron system with exciton mass $m_X= m_e +m_h$.

For each scattering state the coefficients $\alpha(k)$ and $\beta(k)$ are determined from a fit of Eq.~\eqref{eqn:free_particle_2d} to the numerical solution. Performing this fitting procedure separately for symmetric ($S$) and antisymmetric ($A$) states allows us to extract the energy-dependent $s$-wave phase shifts $\delta^{S/A}(k) =\text{arctan}[-\beta(k)/\alpha(k)]$. The resulting phase shifts for MoS$_2$ are depicted in Fig.~\ref{fig:phaseshifts_mobile} as green squares. The energy range $0\leq E\leq 60 \text{meV}$ is chosen to be comparable to typical  Fermi energies (measured from the conduction band edge) realized in gate-doped TMD heterostructures \cite{Sidler2016,Courtade_Chraged_excitons_2017,Back2017}. 

To put these results into context, we compare them to the universal low-energy behavior of $s$-wave phase shifts in two-dimensional systems \cite{verhaar_scattering_1984, adhikari_quantum_1986} 
\begin{equation}
\text{cot} \delta \approx \pi^{-1} \ln{\frac{E}{E_1}},
\label{eqn:delta_lowe}
\end{equation}
where the energy scale $E_1=\hbar^2/2M_\text{red} a_\text{2D}^2$ defines the two-dimensional scattering length $a_\text{2D}$. If a weakly bound state (i.e. a trion) exists in the spectrum close to the exciton-electron scattering threshold, $E_1$ agrees with its binding energy.  Importantly, this holds, however, only as long as its energy remains much smaller than the energy scale $\varepsilon_R=\hbar^2/M_\text{red} r_0^2$ given by the range $r_0$ of interactions. In particular, for zero-range, i.e. contact interactions, Eq.~\eqref{eqn:delta_lowe} becomes exact.

In Fig.~\ref{fig:phaseshifts_mobile} we show the zero-range phase shifts  (blue dashed lines) as obtained from Eq. (\ref{eqn:delta_lowe}) for the symmetric and antisymmetric channel. For symmetric states we show the results for the parameter $E_1$ taken as the trion energy obtained from our numerical calculation ($E_1= E_T=31.7  \text{ meV}$). While the phase shift at low momenta (inset in Fig.~\ref{fig:phaseshifts_mobile}(a)) is well described by Eq.~\eqref{eqn:delta_lowe}, already for energies larger than 1 meV substantial deviations become apparent. For the antisymmetric channel in Fig.~\ref{fig:phaseshifts_mobile}(b), these deviations are even more pronounced. This comparison between the zero-range phase shifts and the numerical result (green symbols) clearly shows that contact interaction models can provide only a rather crude approximation for exciton-electron scattering in TMD. \\

\textbf{Effective exciton-electron scattering model.---} The previous analysis reveals that for a reliable description  of interacting Bose-Fermi mixtures in 2D semiconductors composed of excitons, electrons, and trions  effective low-energy models are required that go beyond contact interactions. Ideally such a low-energy description should capture not only the relevant universal physics but also remain sufficiently simple to be viable as an input for efficient many-body calculations, for instance, using diagramatics \cite{Shelykh2010,Cotlet2016}, QMC \cite{Astrakharchik2004,Boninsegni2006}, variational \cite{Sidler2016}, or field theoretical approaches \cite{Ludwig2011}.

We find that the model
\begin{equation}
\displaystyle
V^{(S/A)}_\text{Xe}(r)= \begin{cases}
V_0&\text{if } r \leq r^*\\
-\frac{\alpha}{2} \left(\frac{d V_K(r)}{d r} \right)^2 &\text{else} .
\end{cases}
\label{eqn:pseudopotential}
\end{equation}
provides such an accurate description of the effective low-energy X-electron scattering in the symmetric ($S$) or antisymmetric ($A$) interaction channel. Here the short-range physics is determined by the depth $V_0$ and the length scale $r^*$ characteristic for the channel $S$ or $A$. The long-range part in turn is fully determined by the polarizability $\alpha$ of the exciton which we calculate from first principles in our approach, see Tab.~\ref{tab:table2}.

The form of Eq.~\eqref{eqn:pseudopotential} can be understood from the observation that the scattering of an electron and an exciton is governed by the electrostatic interaction of a charge (the electron) and a neutral polarizable object (the exciton).
First, the presence of the charge of the scattered electron induces an electric dipole moment $\vec{d}=\alpha \vec{E}$ with $|\vec{E}|\sim dV_K/dr$. Since the energy of an electric dipole $\vec{d}$ in a field $\vec{E}$ scales as $\sim \vec{d} \cdot \vec{E}$, the form of   Eq.~\eqref{eqn:pseudopotential} follows. Naturally, at large separation $r$, or in absence of dielectric screening, one recovers the familiar scaling $V_\text{Xe}(r) \to - \alpha/(2r^4)$ of charge-induced dipole interactions. Note that in the context of quantum chemistry, potentials similar to Eq.~\eqref{eqn:pseudopotential} are successfully applied in the description of the scattering of electrons with charge-neutral atoms \cite{Idziaszek2009} which leads to the electron-mediated binding of Rydberg molecules \cite{greene_creation_2000, bendkowsky_observation_2009}.

 \begin{table}[b]
\begin{tabular}{|l|l|l|l|l|} \hline  & MoS$_2$ & MoSe$_2$& WS$_2$&WSe$_2$ \\ \hline 
{{\bf symmetric scattering}}& & & &\\
$r^*$ (\AA) &  34&   33 & 43&42\\
$V_0$ (meV)& -58.5 &  -51.7  & -60.2&-52.9\\ \hline
{\bf antisymmetric scattering}& & & &\\
$r^*$ (\AA) & 0.265 & 0.26  &0.35 & 0.36 \\ \hline
{\bf Exciton polarizability}& & & &\\
$\alpha$ ($10^3$ a.u.) &52& 61 & 69& 88 \\ \hline 
\end{tabular}
\caption{Parameters $V_0$, $r^*$ and $\alpha$ for the model potential in Eq.~(\ref{eqn:pseudopotential}) that reproduce the exciton-electron $s$-wave phase shifts (symmetric or antisymmetric) for the different TMD materials specified in Table \ref{tab:table1}. We employ a hardcore short-range cutoff, i.e. $V_0\to \infty$, for the antisymmetric channel, while we use finite-depth potentials for the symmetric channel that reproduce also the trion binding energy.  The a.u. of polarizability is $10^{-22} \text{eV}/(\text{m}/\text{V})^2$.}
\label{tab:table2}
\end{table}

To which extent the scattering of excitons and electrons  takes place in the symmetric or in the antisymmetric scattering channel is closely related to the spin and valley degrees of freedom of both electrons. In typical experiments, electrons have a well-defined valley and spin  index (for instance in the case depicted in Fig.~\ref{fig:coordinates}(b) one has $S_1=\{K,\uparrow\}$, $S_2=\{K',\downarrow\}$). Consequently, the scattered electrons have, in general, to be considered as being in a superposition of singlet and triplet scattering states and the exciton-electron interaction is expressed as $V_\text{Xe}(r)= V^S_\text{Xe}(r) \hat{P}_S + V^A_\text{Xe}(r)(1-\hat{P}_S)$ with $\hat{P}_S$  the projector onto the electron-spin singlet channel. This implies that exciton-electron collisions can effectively induce spin flips, in the sense that an initially free $\uparrow$-electron in the $K'$ valley may form --- after the collision with the K-intravalley exciton --- a bound intervalley exciton with the hole in the $K$ valley, leaving behind the formerly bound $\downarrow$-electron in the $K$ valley in a scattering state, given that energy and momentum conservation are satisfied. Additional Coulomb exchange can modify this process by opening or closing scattering channels, since it lifts the energetic degeneracy of intra- and intervalley configurations \cite{Glazov_2015, Plechinger_2016,Courtade_Chraged_excitons_2017}.

In Fig.~\ref{fig:phaseshifts_mobile} we compare phase shift for MoS$_2$ obtained by diagonalization to the result based on the model potential Eq.~\eqref{eqn:pseudopotential} for the parameters $V_0$, $r^*$ and $\alpha$ given in Table \ref{tab:table2}. Here  the polarizabilities $\alpha$ are directly obtained from first principles by a calculation of the quadratic stark spectra of excitons in homogenous electric field via our diagonalization technique. This approach to obtain $\alpha$ is similar to the analysis \cite{Efimkin_Many-body_2017} and yields polarizabilities consistent with \cite{Pedersen_Exciton_Stark_shift_2016}. The parameters $V_0$ and $r^*$ are obtained by fitting the numerically calculated phase shifts.

In fact, we find that one already achieves good agreement for the scattering phase shift when using a hard wall barrier at short-distances, i.e. $V_0\to \infty$, see solid orange line in Fig. \ref{fig:phaseshifts_mobile}~(b). This implies that even simple single-parameter models are sufficient to obtain a reliable description of the exciton-electron scattering above threshold. Using the finite depth $V_0$ as further parameter, one is additionally  able to accurately reproduce the trion binding energy, cf.~the  solid, orange line in Fig.~\ref{fig:phaseshifts_mobile}(a). We note that the scattering phase shifts can not be accurately described by pure hard-sphere potentials for which the phase shifts $\cot \delta= Y_0(k r^*)/J_0(k r^*)$ are analytically known \cite{verhaar_scattering_1984} and ignore the long-range tail of interactions. 

While we show explicit results for $\delta(E)$ only for MoS$_2$, we also provide the effective model parameters for other  TMD materials in Table \ref{tab:table2}. In all cases  we find that accounting for the long-range polarization potential is essential to obtain a reliable low-energy scattering model. 

\section{Optical Absorption spectrum of charge-doped $\text{MoSe}_2$}\label{Sec_ElectronDefectScatt}

By featuring excitons that remain well-defined particles even under substantial electron doping, TMDs allow one to study the regime where a low density of excitons is immersed in a bath of electrons \cite{Mak2012,Sidler2016, Efimkin_Many-body_2017, Efimkin_Exciton-polarons_2018}. This represents a realization of the many-body problem of Fermi polarons, where one considers the interaction of a single mobile impurity (the exciton) with a sea of fermions (electrons) \cite{Chevy2006,Prokofev_Svistunov_PRB_2008, Punk2009, Schmidt_Enss_PRA_2011,Parish_Meera_PRA_2013,Massignan2014}.  The dressing of the impurity by particle-hole excitations of the Fermi bath leads to renormalized properties of the impurity which becomes a quasiparticle, the Fermi polaron \cite{Rosch_1999, Chevy2006, Prokofev_Svistunov_PRB_2008}. Key signatures of polaron formation are a renormalized mass, a reduced  absorption line strength, and a shift of the impurity energy. The formation of attractive and repulsive Fermi polaron branches has been predicted for two-dimensional systems in \cite{Schmidt2012b} and were observed first in the radio-frequency response of ultracold atomic gases \cite{Koschorreck2012}.

Recently, signatures of Fermi polarons in TMDs were reported in \cite{Sidler2016} where  polaron energy shifts were measured in gate-tunable monolayer MoSe$_2$ for variable Fermi energies $\epsilon_F$ in the range of $0 \leq \epsilon_F \leq 40 \text{ meV}$. As predicted and observed in the context of ultracold atoms, also in TMDs two polaron branches exist. The so-called attractive polaron branch corresponds to the exciton being dressed by the virtual occupation of the trion state in addition to particle-hole excitations of the Fermi sea. In the limit of low charge-carrier density this branch emerges in the absorption spectrum at the trion energy. In contrast, the so-called repulsive polaron corresponds to the exciton being dressed predominately by particle-hole excitations of the Fermi sea. This leads to a repulsive blue shift of the bare exciton line as more charge carriers are inserted into the system.

Previous analysis compared the experimentally observed absorption spectrum with a many-body model that assumes contact interactions between electrons and excitons \cite{Sidler2016,Efimkin_Many-body_2017,Efimkin_Exciton-polarons_2018}.
Under this assumption experimental absorption line shifts were found to be in relatively good agreement with a variational calculation that takes into account the dressing of the exciton by single particle-hole excitations of the Fermi sea \cite{Chevy2006}. 

While theory and experiment agree well at low charge-carrier doping, at higher Fermi energies increasing deviations are found in particular for the attractive polaron branch when $\epsilon_F>20$ meV \cite{Sidler2016}. One possible source for discrepancies, which is strongly suggested by our analysis, are finite-range corrections of the exciton-electron interaction. Our results for the phase shifts in Sec.~\ref{Sec_ElectronXScatt} show that these corrections become increasingly important at larger Fermi energies.

In order to estimate the role and extent of the finite-range corrections on polaron energies we apply here Fumi's theorem, which links the energy shift of absorption lines to the phase shifts of the electron-impurity interactions via \cite{fumi_1955,schmidt2017,Pimenov_2018}  
\begin{align}
E_\text{pol}^\text{rep} (\epsilon_F) &= -\int \limits_0^{\epsilon_F} \frac{dE}{\pi} \delta(E)  \nonumber \\
E_\text{pol}^\text{att} (\epsilon_F) &= E_\text{Tr}- \epsilon_F-\int \limits_0^{\epsilon_F} \frac{dE}{\pi} \delta(E).
\label{eqn:fumi}
\end{align}
Here the Fermi-energy $\epsilon_F$ is measured from the conduction band edge, $E_\text{Tr}$ is the trion binding energy, and $\delta(E)$ is the exciton-electron phase shift. For simplicity and in order to allow comparison to \cite{Sidler2016,Efimkin_Many-body_2017}, the spin dependence of the phase shifts is neglected. Hence interactions take place exclusively in the single scattering channel that supports the trion state. Fumi's theorem becomes exact for an infinite impurity mass, while for the present case it is an approximation that neglects recoil but includes infinitely many particle-hole excitations of the Fermi sea.

\begin{figure}[t]
\includegraphics[width= 1 \linewidth]{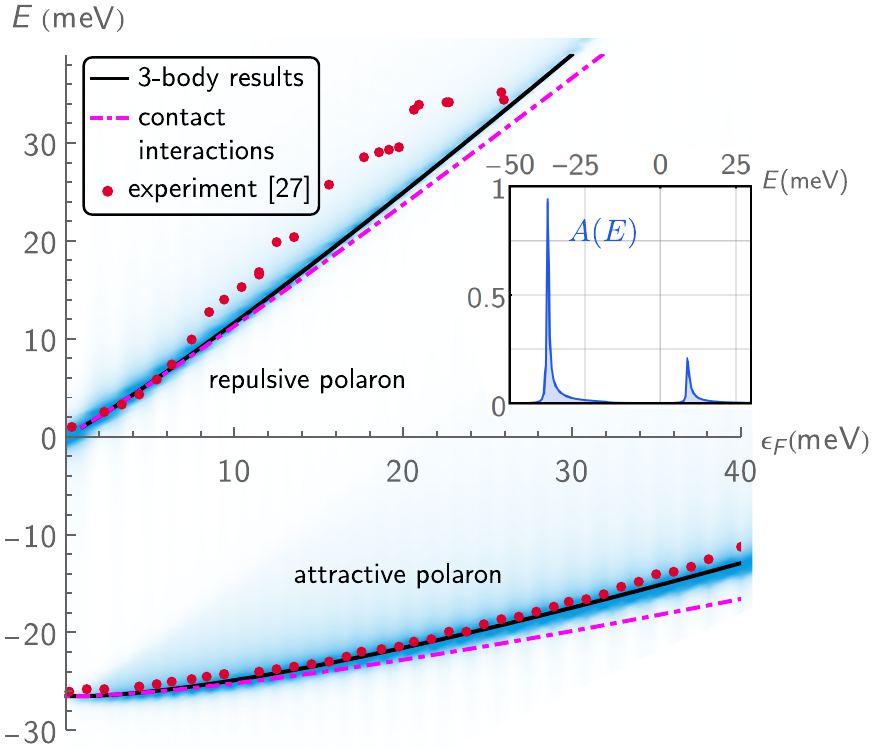}
\caption{Energies of the attractive and repulsive polaron for a single exciton impurity in electron-doped monolayer MoSe$_2$ as a function of the Fermi energy $\epsilon_F$.
We compare theoretical predictions  based on Fumi's theorem (\ref{eqn:fumi}) and employ exciton-electron phase shifts resulting either from our three-body simulations (solid black lines) or from contact interactions (dashed-dotted magenta lines). Deviations of these predictions characterize the impact of finite-range corrections on the polaron. Information on the lineshape and line strength of the polaron resonances is provided by the simulated polaron absorption spectrum $A(E)$ (blue shading). We obtain $A(E)$ employing the FDA and the model potential Eq.~(\ref{eqn:pseudopotential}) that accounts for finite-range corrections. The inset shows a cut of $A(E)$ for fixed doping $\epsilon_F=20$~meV. Optical absorption measurements \cite{Sidler2016} in gate-tuneable monolayer MoSe$_2$ (red dots) are shown for comparison. All theoretical energies are shifted by $0.8 \, \epsilon_F$.}
\label{fig:sidler}
\end{figure}

Fig.~\ref{fig:sidler} shows a comparison of polaron energies obtained from Eq.~\eqref{eqn:fumi} using either the complete three-body result $\delta^{S}(E)$ (solid black line) as input or the contact interaction Eq.~(\ref{eqn:delta_lowe}) (dashed-dotted magenta line).
The impact of finite-range corrections is visible in the difference between the two predictions. While both energies coincide at low Fermi energies $\epsilon_F$, there are increasing deviations for larger $\epsilon_F$ with the zero-range approximation systematically underestimating the polaron energy. For instance for $\epsilon_F= 40 \text{ meV}$, deviations are on the order of 5 meV. Note, for a comparison with the experiment and following \cite{Sidler2016} all theoretical energies are shifted by $0.8 \,\epsilon_F$, see below.

In addition to polaron energies one can also predict the spectral lineshape using the model potential (\ref{eqn:pseudopotential}) and the functional determinant approach (FDA); for details see App.~\ref{sec:appendix_FDA}. The resulting spectrum $A(E)$ is shown in Fig.~\ref{fig:sidler} as blue shading together with a cut of $A(E)$ for $\epsilon_F=$~20~meV in the inset. Similar to Fumi's theorem the FDA neglects impurity recoil but includes finite-range effects of the underlying interaction as well as infinitely many particle-hole excitations. As a consequence, peak positions of $A(E)$ coincide with the energies predicted by Eq.~\eqref{eqn:fumi} for the numerically exact phase shifts (solid black line). While the signal strength of the repulsive polaron decreases and broadens rapidly as a function of $\epsilon_F$, the visibility of the attractive polaron increases, in good agreement with alternative theoretical models \cite{Sidler2016, Efimkin_Many-body_2017}. 

For comparison we present in Fig.~\ref{fig:sidler} also polaron energies experimentally obtained in Ref.~\cite{Sidler2016} (red dots). The measured energy positions agree overall well with our results. Furthermore the larger spread of experimental data for the repulsive polaron at $\epsilon_F \gtrsim 8$~ meV is consistent with the reduced signal strength predicted by the FDA. As noted above all theoretical energies are shifted by the linear function $0.8\, \epsilon_F$ which was determined in Ref.~\cite{Sidler2016} to account for the combined effects of phase space filling, screening and band gap renormalization \cite{Tiene_Levinsen_Parish_MacDonald_Keeling_Marchetti_2019}.  

In conclusion our analysis based on Fumi's theorem and the FDA demonstrates that finite-range corrections have indeed a significant impact on polaron energies at typical charge-carrier densities and need to be taken into account for a correct description of the corresponding optical response. Furthermore, our results provide an illustrative example of how the exciton-electron phase shifts and the interaction potentials derived in this work can be incorporated into many-body theories to account for a detailed description of electron-electron scattering.  

We note that more accurate descriptions of polaron spectra in TMDs should consider also the impurity recoil as well as phase-space filling, band gap renormalization and screening. Moreover, scattering between the $K$-valley exciton and electrons in the same valley can become relevant which we neglected here for simplicity and to allow for comparison with the theoretical model in Refs.~\cite{Sidler2016, Efimkin_Many-body_2017}. This scattering will provide an additional interaction channel contributing to Fumi's theorem by the scattering phase shift $\delta^{A}$  shown in Fig.~\ref{fig:phaseshifts_mobile}(b), and thus will lead to an additional, yet smaller, energy shift of the polaron. Finally, as the Fermi energy exceeds the spin-orbit splitting $\Delta_\text{SOC}^c$ of the conduction bands, see Fig.~\ref{fig:coordinates}, additional charge carriers start to interact with the exciton and will contribute to further shifts of the polaron lines in absorption and photoluminescence experiments.

\section{Conclusions}
\label{SecSummary}

In this work we introduced an exact, discrete variable representation-based diagonalization approach  to the scattering of electrons and holes in two-dimensional transition metal dicalchogenides. We predicted the trion and exciton binding energies which are in exellent agreement with QMC predictions \cite{kylanpaa_binding_2015}, and that can serve as benchmarks for variational approaches \cite{semina2008,Courtade_Chraged_excitons_2017,wang2018}. The diagonalization yields also the spectrum and wave functions of excited states which makes the approach an alternative to other methods such as Faddeev equation formalism \cite{filikhin2018trions}, variational optimization \cite{berkelbach_theory_2013, Courtade_Chraged_excitons_2017}, diagrammatics \cite{Combescot2011}, or path-integral and diffusion Monte Carlo \cite{kylanpaa_binding_2015} . The excited states fall into two categories: bound excited trions and spatially extended states that correspond to electrons scattering with tightly bound excitons.  Using asymptotic wave functions we predicted the energy-dependent scattering phase shifts which determine the strength of exciton-electron interactions in TMDs.

Our investigation shows that contact interaction models are insufficient to describe these phase shifts. However, still relatively simple scattering models can be derived, and we introduce a model potential that can accurately capture the predicted scattering phase shifts over a large range of energy scales. The potential may be used as a reliable input for many-body models of Bose-Fermi mixtures consisting of electrons and excitons in TMDs and we provide numerical parameters for monolayers of MoS$_2$, MoSe$_2$, WS$_2$, and WSe$_2$. As an application, we showed that the prediction for the optical absorption spectra of n-doped MoSe$_2$ based  on the potential,  yields good agreement with recent experiments that explored Fermi polaron formation \cite{Sidler2016,Ravets2018}.

In this work we focussed on introducing the exact diagonalization approach and on providing a benchmark of exact diagonalization against QMC. Hence, various effects were not included that  yield further quantitative corrections, such as the effect of Berry curvature, band warping, or electron-hole exchange. All of these effects can in principle be included in our Hamiltonian approach. Moreover, we employed the simple Keldysh potential Eq.~\eqref{eqn:Keldysh potential} to describe the interactions between charge carriers. From the field solutions of the Laplace equation with appropriate boundary conditions more accurate potentials can be derived that account for a finite material thickness and the presence of various dielectric environments \cite{florian2018}. Our approach can be extended to include magnetic fields \cite{Stier_2018, Goryca_Stier_2019} and defects \cite{Klein_2019} and it provides means to explore multilayer heterostructures that hold promise to exhibit striking many-body physics. In particular, depending on material stacking, exciton and trion energies can be manipulated and scattering phase shifts may exhibit strong-coupling behavior. 

We finally emphasize that the exact diagonalization approach is generally applicable to  systems where bound states emerge in the scattering of multiple particles. Our approach allows one to generically derive effective low-energy models for the description of the many-body physics and scattering of the resulting composite objects, ranging from dimer formation in cold atomic systems \cite{Petrov2004} to dephasing in storage-of-light experiments induced by multi-body Rydberg molecules  \cite{greene_creation_2000,Fey2019,Mirgorodskiy2017}.
\vspace{5pt}

\subsection*{Acknowledgments}
We thank H.D. Meyer for providing helpful information concerning the DVR method. C.~F. gratefully acknowledges support by the Studienstiftung des deutschen Volkes and the hospitality of the Institute for Theoretical Atomic, Molecular and Optical Physics (ITAMP). R.~S. is supported by the Deutsche Forschungsgemeinschaft (DFG, German Research Foundation) under Germany's Excellence Strategy -- EXC-2111 -- project ID 390814868. P.~S. acknowledges the support by the Cluster of Excellence `Advanced Imaging of Matter' of the Deutsche Forschungsgemeinschaft (DFG, German Research Foundation) --EXC-2056-- project ID 390715994. A.~I. is supported by
a European Research Council (ERC) Advanced investigator
grant (POLTDES).


\begin{thebibliography}{95}%
\makeatletter
\providecommand \@ifxundefined [1]{%
 \@ifx{#1\undefined}
}%
\providecommand \@ifnum [1]{%
 \ifnum #1\expandafter \@firstoftwo
 \else \expandafter \@secondoftwo
 \fi
}%
\providecommand \@ifx [1]{%
 \ifx #1\expandafter \@firstoftwo
 \else \expandafter \@secondoftwo
 \fi
}%
\providecommand \natexlab [1]{#1}%
\providecommand \enquote  [1]{``#1''}%
\providecommand \bibnamefont  [1]{#1}%
\providecommand \bibfnamefont [1]{#1}%
\providecommand \citenamefont [1]{#1}%
\providecommand \href@noop [0]{\@secondoftwo}%
\providecommand \href [0]{\begingroup \@sanitize@url \@href}%
\providecommand \@href[1]{\@@startlink{#1}\@@href}%
\providecommand \@@href[1]{\endgroup#1\@@endlink}%
\providecommand \@sanitize@url [0]{\catcode `\\12\catcode `\$12\catcode
  `\&12\catcode `\#12\catcode `\^12\catcode `\_12\catcode `\%12\relax}%
\providecommand \@@startlink[1]{}%
\providecommand \@@endlink[0]{}%
\providecommand \url  [0]{\begingroup\@sanitize@url \@url }%
\providecommand \@url [1]{\endgroup\@href {#1}{\urlprefix }}%
\providecommand \urlprefix  [0]{URL }%
\providecommand \Eprint [0]{\href }%
\providecommand \doibase [0]{http://dx.doi.org/}%
\providecommand \selectlanguage [0]{\@gobble}%
\providecommand \bibinfo  [0]{\@secondoftwo}%
\providecommand \bibfield  [0]{\@secondoftwo}%
\providecommand \translation [1]{[#1]}%
\providecommand \BibitemOpen [0]{}%
\providecommand \bibitemStop [0]{}%
\providecommand \bibitemNoStop [0]{.\EOS\space}%
\providecommand \EOS [0]{\spacefactor3000\relax}%
\providecommand \BibitemShut  [1]{\csname bibitem#1\endcsname}%
\let\auto@bib@innerbib\@empty
\bibitem [{\citenamefont {Bloch}\ \emph {et~al.}(2008)\citenamefont {Bloch},
  \citenamefont {Dalibard},\ and\ \citenamefont {Zwerger}}]{Bloch2008}%
  \BibitemOpen
  \bibfield  {author} {\bibinfo {author} {\bibfnamefont {I.}~\bibnamefont
  {Bloch}}, \bibinfo {author} {\bibfnamefont {J.}~\bibnamefont {Dalibard}}, \
  and\ \bibinfo {author} {\bibfnamefont {W.}~\bibnamefont {Zwerger}},\ }\href
  {\doibase 10.1103/RevModPhys.80.885} {\bibfield  {journal} {\bibinfo
  {journal} {Rev. Mod. Phys.}\ }\textbf {\bibinfo {volume} {80}},\ \bibinfo
  {pages} {885} (\bibinfo {year} {2008})}\BibitemShut {NoStop}%
\bibitem [{\citenamefont {Lous}\ \emph {et~al.}(2018)\citenamefont {Lous},
  \citenamefont {Fritsche}, \citenamefont {Jag}, \citenamefont {Lehmann},
  \citenamefont {Kirilov}, \citenamefont {Huang},\ and\ \citenamefont
  {Grimm}}]{Lous2018}%
  \BibitemOpen
  \bibfield  {author} {\bibinfo {author} {\bibfnamefont {R.~S.}\ \bibnamefont
  {Lous}}, \bibinfo {author} {\bibfnamefont {I.}~\bibnamefont {Fritsche}},
  \bibinfo {author} {\bibfnamefont {M.}~\bibnamefont {Jag}}, \bibinfo {author}
  {\bibfnamefont {F.}~\bibnamefont {Lehmann}}, \bibinfo {author} {\bibfnamefont
  {E.}~\bibnamefont {Kirilov}}, \bibinfo {author} {\bibfnamefont
  {B.}~\bibnamefont {Huang}}, \ and\ \bibinfo {author} {\bibfnamefont
  {R.}~\bibnamefont {Grimm}},\ }\href {\doibase 10.1103/PhysRevLett.120.243403}
  {\bibfield  {journal} {\bibinfo  {journal} {Phys. Rev. Lett.}\ }\textbf
  {\bibinfo {volume} {120}},\ \bibinfo {pages} {243403} (\bibinfo {year}
  {2018})}\BibitemShut {NoStop}%
\bibitem [{\citenamefont {Fr\"ohlich}(1950)}]{Froh3}%
  \BibitemOpen
  \bibfield  {author} {\bibinfo {author} {\bibfnamefont {H.}~\bibnamefont
  {Fr\"ohlich}},\ }\href {\doibase 10.1103/PhysRev.79.845} {\bibfield
  {journal} {\bibinfo  {journal} {Phys. Rev.}\ }\textbf {\bibinfo {volume}
  {79}},\ \bibinfo {pages} {845} (\bibinfo {year} {1950})}\BibitemShut
  {NoStop}%
\bibitem [{\citenamefont {Fr\"ohlich}(1952)}]{Froh4}%
  \BibitemOpen
  \bibfield  {author} {\bibinfo {author} {\bibfnamefont {H.}~\bibnamefont
  {Fr\"ohlich}},\ }\href {\doibase 10.1098/rspa.1952.0212} {\bibfield
  {journal} {\bibinfo  {journal} {Proc. R. Soc. A}\ }\textbf {\bibinfo {volume}
  {215}},\ \bibinfo {pages} {291} (\bibinfo {year} {1952})}\BibitemShut
  {NoStop}%
\bibitem [{\citenamefont {Rath}\ and\ \citenamefont
  {Schmidt}(2013)}]{Rath2013}%
  \BibitemOpen
  \bibfield  {author} {\bibinfo {author} {\bibfnamefont {S.~P.}\ \bibnamefont
  {Rath}}\ and\ \bibinfo {author} {\bibfnamefont {R.}~\bibnamefont {Schmidt}},\
  }\href {\doibase 10.1103/PhysRevA.88.053632} {\bibfield  {journal} {\bibinfo
  {journal} {Phys. Rev. A}\ }\textbf {\bibinfo {volume} {88}},\ \bibinfo
  {pages} {053632} (\bibinfo {year} {2013})}\BibitemShut {NoStop}%
\bibitem [{\citenamefont {Hu}\ \emph {et~al.}(2016)\citenamefont {Hu},
  \citenamefont {Van~de Graaff}, \citenamefont {Kedar}, \citenamefont {Corson},
  \citenamefont {Cornell},\ and\ \citenamefont {Jin}}]{Hu2016}%
  \BibitemOpen
  \bibfield  {author} {\bibinfo {author} {\bibfnamefont {M.-G.}\ \bibnamefont
  {Hu}}, \bibinfo {author} {\bibfnamefont {M.~J.}\ \bibnamefont {Van~de
  Graaff}}, \bibinfo {author} {\bibfnamefont {D.}~\bibnamefont {Kedar}},
  \bibinfo {author} {\bibfnamefont {J.~P.}\ \bibnamefont {Corson}}, \bibinfo
  {author} {\bibfnamefont {E.~A.}\ \bibnamefont {Cornell}}, \ and\ \bibinfo
  {author} {\bibfnamefont {D.~S.}\ \bibnamefont {Jin}},\ }\href {\doibase
  10.1103/PhysRevLett.117.055301} {\bibfield  {journal} {\bibinfo  {journal}
  {Phys. Rev. Lett.}\ }\textbf {\bibinfo {volume} {117}},\ \bibinfo {pages}
  {055301} (\bibinfo {year} {2016})}\BibitemShut {NoStop}%
\bibitem [{\citenamefont {J\o{}rgensen}\ \emph {et~al.}(2016)\citenamefont
  {J\o{}rgensen}, \citenamefont {Wacker}, \citenamefont {Skalmstang},
  \citenamefont {Parish}, \citenamefont {Levinsen}, \citenamefont
  {Christensen}, \citenamefont {Bruun},\ and\ \citenamefont
  {Arlt}}]{Jorgensen2016}%
  \BibitemOpen
  \bibfield  {author} {\bibinfo {author} {\bibfnamefont {N.~B.}\ \bibnamefont
  {J\o{}rgensen}}, \bibinfo {author} {\bibfnamefont {L.}~\bibnamefont
  {Wacker}}, \bibinfo {author} {\bibfnamefont {K.~T.}\ \bibnamefont
  {Skalmstang}}, \bibinfo {author} {\bibfnamefont {M.~M.}\ \bibnamefont
  {Parish}}, \bibinfo {author} {\bibfnamefont {J.}~\bibnamefont {Levinsen}},
  \bibinfo {author} {\bibfnamefont {R.~S.}\ \bibnamefont {Christensen}},
  \bibinfo {author} {\bibfnamefont {G.~M.}\ \bibnamefont {Bruun}}, \ and\
  \bibinfo {author} {\bibfnamefont {J.~J.}\ \bibnamefont {Arlt}},\ }\href
  {\doibase 10.1103/PhysRevLett.117.055302} {\bibfield  {journal} {\bibinfo
  {journal} {Phys. Rev. Lett.}\ }\textbf {\bibinfo {volume} {117}},\ \bibinfo
  {pages} {055302} (\bibinfo {year} {2016})}\BibitemShut {NoStop}%
\bibitem [{\citenamefont {Camargo}\ \emph {et~al.}(2018)\citenamefont
  {Camargo}, \citenamefont {Schmidt}, \citenamefont {Whalen}, \citenamefont
  {Ding}, \citenamefont {Woehl}, \citenamefont {Yoshida}, \citenamefont
  {Burgd\"orfer}, \citenamefont {Dunning}, \citenamefont {Sadeghpour},
  \citenamefont {Demler},\ and\ \citenamefont {Killian}}]{Camargo2018}%
  \BibitemOpen
  \bibfield  {author} {\bibinfo {author} {\bibfnamefont {F.}~\bibnamefont
  {Camargo}}, \bibinfo {author} {\bibfnamefont {R.}~\bibnamefont {Schmidt}},
  \bibinfo {author} {\bibfnamefont {J.~D.}\ \bibnamefont {Whalen}}, \bibinfo
  {author} {\bibfnamefont {R.}~\bibnamefont {Ding}}, \bibinfo {author}
  {\bibfnamefont {G.}~\bibnamefont {Woehl}}, \bibinfo {author} {\bibfnamefont
  {S.}~\bibnamefont {Yoshida}}, \bibinfo {author} {\bibfnamefont
  {J.}~\bibnamefont {Burgd\"orfer}}, \bibinfo {author} {\bibfnamefont {F.~B.}\
  \bibnamefont {Dunning}}, \bibinfo {author} {\bibfnamefont {H.~R.}\
  \bibnamefont {Sadeghpour}}, \bibinfo {author} {\bibfnamefont
  {E.}~\bibnamefont {Demler}}, \ and\ \bibinfo {author} {\bibfnamefont {T.~C.}\
  \bibnamefont {Killian}},\ }\href {\doibase 10.1103/PhysRevLett.120.083401}
  {\bibfield  {journal} {\bibinfo  {journal} {Phys. Rev. Lett.}\ }\textbf
  {\bibinfo {volume} {120}},\ \bibinfo {pages} {083401} (\bibinfo {year}
  {2018})}\BibitemShut {NoStop}%
\bibitem [{\citenamefont {Imamoglu}\ \emph {et~al.}(1996)\citenamefont
  {Imamoglu}, \citenamefont {Ram}, \citenamefont {Pau},\ and\ \citenamefont
  {Yamamoto}}]{Imamoglu1996}%
  \BibitemOpen
  \bibfield  {author} {\bibinfo {author} {\bibfnamefont {A.}~\bibnamefont
  {Imamoglu}}, \bibinfo {author} {\bibfnamefont {R.~J.}\ \bibnamefont {Ram}},
  \bibinfo {author} {\bibfnamefont {S.}~\bibnamefont {Pau}}, \ and\ \bibinfo
  {author} {\bibfnamefont {Y.}~\bibnamefont {Yamamoto}},\ }\href {\doibase
  10.1103/PhysRevA.53.4250} {\bibfield  {journal} {\bibinfo  {journal} {Phys.
  Rev. A}\ }\textbf {\bibinfo {volume} {53}},\ \bibinfo {pages} {4250}
  (\bibinfo {year} {1996})}\BibitemShut {NoStop}%
\bibitem [{\citenamefont {Kasprzak}\ \emph {et~al.}(2006)\citenamefont
  {Kasprzak}, \citenamefont {Richard}, \citenamefont {Kundermann},
  \citenamefont {Baas}, \citenamefont {Jeambrun}, \citenamefont {Keeling},
  \citenamefont {Marchetti}, \citenamefont {Szyma{\'n}ska}, \citenamefont
  {Andr{\'e}}, \citenamefont {Staehli}, \citenamefont {Savona}, \citenamefont
  {Littlewood}, \citenamefont {Deveaud},\ and\ \citenamefont
  {Dang}}]{Kasprzak2006}%
  \BibitemOpen
  \bibfield  {author} {\bibinfo {author} {\bibfnamefont {J.}~\bibnamefont
  {Kasprzak}}, \bibinfo {author} {\bibfnamefont {M.}~\bibnamefont {Richard}},
  \bibinfo {author} {\bibfnamefont {S.}~\bibnamefont {Kundermann}}, \bibinfo
  {author} {\bibfnamefont {A.}~\bibnamefont {Baas}}, \bibinfo {author}
  {\bibfnamefont {P.}~\bibnamefont {Jeambrun}}, \bibinfo {author}
  {\bibfnamefont {J.~M.~J.}\ \bibnamefont {Keeling}}, \bibinfo {author}
  {\bibfnamefont {F.~M.}\ \bibnamefont {Marchetti}}, \bibinfo {author}
  {\bibfnamefont {M.~H.}\ \bibnamefont {Szyma{\'n}ska}}, \bibinfo {author}
  {\bibfnamefont {R.}~\bibnamefont {Andr{\'e}}}, \bibinfo {author}
  {\bibfnamefont {J.~L.}\ \bibnamefont {Staehli}}, \bibinfo {author}
  {\bibfnamefont {V.}~\bibnamefont {Savona}}, \bibinfo {author} {\bibfnamefont
  {P.~B.}\ \bibnamefont {Littlewood}}, \bibinfo {author} {\bibfnamefont
  {B.}~\bibnamefont {Deveaud}}, \ and\ \bibinfo {author} {\bibfnamefont
  {L.~S.}\ \bibnamefont {Dang}},\ }\href {\doibase 10.1038/nature05131}
  {\bibfield  {journal} {\bibinfo  {journal} {Nature}\ }\textbf {\bibinfo
  {volume} {443}},\ \bibinfo {pages} {409} (\bibinfo {year}
  {2006})}\BibitemShut {NoStop}%
\bibitem [{\citenamefont {Carusotto}\ and\ \citenamefont
  {Ciuti}(2013)}]{Carusotto2013}%
  \BibitemOpen
  \bibfield  {author} {\bibinfo {author} {\bibfnamefont {I.}~\bibnamefont
  {Carusotto}}\ and\ \bibinfo {author} {\bibfnamefont {C.}~\bibnamefont
  {Ciuti}},\ }\href {\doibase 10.1103/RevModPhys.85.299} {\bibfield  {journal}
  {\bibinfo  {journal} {Rev. Mod. Phys.}\ }\textbf {\bibinfo {volume} {85}},\
  \bibinfo {pages} {299} (\bibinfo {year} {2013})}\BibitemShut {NoStop}%
\bibitem [{\citenamefont {Deng}\ \emph {et~al.}(2010)\citenamefont {Deng},
  \citenamefont {Haug},\ and\ \citenamefont {Yamamoto}}]{Deng2010}%
  \BibitemOpen
  \bibfield  {author} {\bibinfo {author} {\bibfnamefont {H.}~\bibnamefont
  {Deng}}, \bibinfo {author} {\bibfnamefont {H.}~\bibnamefont {Haug}}, \ and\
  \bibinfo {author} {\bibfnamefont {Y.}~\bibnamefont {Yamamoto}},\ }\href
  {\doibase 10.1103/RevModPhys.82.1489} {\bibfield  {journal} {\bibinfo
  {journal} {Rev. Mod. Phys.}\ }\textbf {\bibinfo {volume} {82}},\ \bibinfo
  {pages} {1489} (\bibinfo {year} {2010})}\BibitemShut {NoStop}%
\bibitem [{\citenamefont {Novoselov}\ \emph {et~al.}(2004)\citenamefont
  {Novoselov}, \citenamefont {Geim}, \citenamefont {Morozov}, \citenamefont
  {Jiang}, \citenamefont {Zhang}, \citenamefont {Dubonos}, \citenamefont
  {Grigorieva},\ and\ \citenamefont {Firsov}}]{Novoselov2004}%
  \BibitemOpen
  \bibfield  {author} {\bibinfo {author} {\bibfnamefont {K.~S.}\ \bibnamefont
  {Novoselov}}, \bibinfo {author} {\bibfnamefont {A.~K.}\ \bibnamefont {Geim}},
  \bibinfo {author} {\bibfnamefont {S.~V.}\ \bibnamefont {Morozov}}, \bibinfo
  {author} {\bibfnamefont {D.}~\bibnamefont {Jiang}}, \bibinfo {author}
  {\bibfnamefont {Y.}~\bibnamefont {Zhang}}, \bibinfo {author} {\bibfnamefont
  {S.~V.}\ \bibnamefont {Dubonos}}, \bibinfo {author} {\bibfnamefont {I.~V.}\
  \bibnamefont {Grigorieva}}, \ and\ \bibinfo {author} {\bibfnamefont {A.~A.}\
  \bibnamefont {Firsov}},\ }\href {\doibase 10.1126/science.1102896} {\bibfield
   {journal} {\bibinfo  {journal} {Science}\ }\textbf {\bibinfo {volume}
  {306}},\ \bibinfo {pages} {666} (\bibinfo {year} {2004})}\BibitemShut
  {NoStop}%
\bibitem [{\citenamefont {Zhang}\ \emph {et~al.}(2005)\citenamefont {Zhang},
  \citenamefont {Tan}, \citenamefont {Stormer},\ and\ \citenamefont
  {Kim}}]{Zhang2005}%
  \BibitemOpen
  \bibfield  {author} {\bibinfo {author} {\bibfnamefont {Y.}~\bibnamefont
  {Zhang}}, \bibinfo {author} {\bibfnamefont {Y.-W.}\ \bibnamefont {Tan}},
  \bibinfo {author} {\bibfnamefont {H.~L.}\ \bibnamefont {Stormer}}, \ and\
  \bibinfo {author} {\bibfnamefont {P.}~\bibnamefont {Kim}},\ }\href {\doibase
  10.1038/nature04235} {\bibfield  {journal} {\bibinfo  {journal} {Nature}\
  }\textbf {\bibinfo {volume} {438}},\ \bibinfo {pages} {201} (\bibinfo {year}
  {2005})}\BibitemShut {NoStop}%
\bibitem [{\citenamefont {Geim}\ and\ \citenamefont
  {Novoselov}(2007)}]{Geim2007}%
  \BibitemOpen
  \bibfield  {author} {\bibinfo {author} {\bibfnamefont {A.~K.}\ \bibnamefont
  {Geim}}\ and\ \bibinfo {author} {\bibfnamefont {K.~S.}\ \bibnamefont
  {Novoselov}},\ }\href {https://doi.org/10.1038/nmat1849} {\bibfield
  {journal} {\bibinfo  {journal} {Nature Mater.}\ }\textbf {\bibinfo {volume}
  {6}},\ \bibinfo {pages} {183} (\bibinfo {year} {2007})}\BibitemShut {NoStop}%
\bibitem [{\citenamefont {Novoselov}\ \emph {et~al.}(2005)\citenamefont
  {Novoselov}, \citenamefont {Jiang}, \citenamefont {Schedin}, \citenamefont
  {Booth}, \citenamefont {Khotkevich}, \citenamefont {Morozov},\ and\
  \citenamefont {Geim}}]{Novoselov2005}%
  \BibitemOpen
  \bibfield  {author} {\bibinfo {author} {\bibfnamefont {K.~S.}\ \bibnamefont
  {Novoselov}}, \bibinfo {author} {\bibfnamefont {D.}~\bibnamefont {Jiang}},
  \bibinfo {author} {\bibfnamefont {F.}~\bibnamefont {Schedin}}, \bibinfo
  {author} {\bibfnamefont {T.~J.}\ \bibnamefont {Booth}}, \bibinfo {author}
  {\bibfnamefont {V.~V.}\ \bibnamefont {Khotkevich}}, \bibinfo {author}
  {\bibfnamefont {S.~V.}\ \bibnamefont {Morozov}}, \ and\ \bibinfo {author}
  {\bibfnamefont {A.~K.}\ \bibnamefont {Geim}},\ }\href {\doibase
  10.1073/pnas.0502848102} {\bibfield  {journal} {\bibinfo  {journal} {Proc.
  Natl. Acad. Sci. U.S.A.}\ }\textbf {\bibinfo {volume} {102}},\ \bibinfo
  {pages} {10451} (\bibinfo {year} {2005})}\BibitemShut {NoStop}%
\bibitem [{\citenamefont {Xi}\ \emph {et~al.}(2015)\citenamefont {Xi},
  \citenamefont {Wang}, \citenamefont {Zhao}, \citenamefont {Park},
  \citenamefont {Law}, \citenamefont {Berger}, \citenamefont {Forr{\'o}},
  \citenamefont {Shan},\ and\ \citenamefont {Mak}}]{Xi2015}%
  \BibitemOpen
  \bibfield  {author} {\bibinfo {author} {\bibfnamefont {X.}~\bibnamefont
  {Xi}}, \bibinfo {author} {\bibfnamefont {Z.}~\bibnamefont {Wang}}, \bibinfo
  {author} {\bibfnamefont {W.}~\bibnamefont {Zhao}}, \bibinfo {author}
  {\bibfnamefont {J.-H.}\ \bibnamefont {Park}}, \bibinfo {author}
  {\bibfnamefont {K.~T.}\ \bibnamefont {Law}}, \bibinfo {author} {\bibfnamefont
  {H.}~\bibnamefont {Berger}}, \bibinfo {author} {\bibfnamefont
  {L.}~\bibnamefont {Forr{\'o}}}, \bibinfo {author} {\bibfnamefont
  {J.}~\bibnamefont {Shan}}, \ and\ \bibinfo {author} {\bibfnamefont {K.~F.}\
  \bibnamefont {Mak}},\ }\href {https://doi.org/10.1038/nphys3538} {\bibfield
  {journal} {\bibinfo  {journal} {Nature Physics}\ }\textbf {\bibinfo {volume}
  {12}},\ \bibinfo {pages} {139} (\bibinfo {year} {2015})}\BibitemShut
  {NoStop}%
\bibitem [{\citenamefont {Bistritzer}\ and\ \citenamefont
  {MacDonald}(2011)}]{Bistritzer2011}%
  \BibitemOpen
  \bibfield  {author} {\bibinfo {author} {\bibfnamefont {R.}~\bibnamefont
  {Bistritzer}}\ and\ \bibinfo {author} {\bibfnamefont {A.~H.}\ \bibnamefont
  {MacDonald}},\ }\href {\doibase 10.1073/pnas.1108174108} {\bibfield
  {journal} {\bibinfo  {journal} {Proc. Natl. Acad. Sci. U.S.A.}\ }\textbf
  {\bibinfo {volume} {108}},\ \bibinfo {pages} {12233} (\bibinfo {year}
  {2011})}\BibitemShut {NoStop}%
\bibitem [{\citenamefont {Cao}\ \emph {et~al.}(2018{\natexlab{a}})\citenamefont
  {Cao}, \citenamefont {Fatemi}, \citenamefont {Fang}, \citenamefont
  {Watanabe}, \citenamefont {Taniguchi}, \citenamefont {Kaxiras},\ and\
  \citenamefont {Jarillo-Herrero}}]{Cao2018}%
  \BibitemOpen
  \bibfield  {author} {\bibinfo {author} {\bibfnamefont {Y.}~\bibnamefont
  {Cao}}, \bibinfo {author} {\bibfnamefont {V.}~\bibnamefont {Fatemi}},
  \bibinfo {author} {\bibfnamefont {S.}~\bibnamefont {Fang}}, \bibinfo {author}
  {\bibfnamefont {K.}~\bibnamefont {Watanabe}}, \bibinfo {author}
  {\bibfnamefont {T.}~\bibnamefont {Taniguchi}}, \bibinfo {author}
  {\bibfnamefont {E.}~\bibnamefont {Kaxiras}}, \ and\ \bibinfo {author}
  {\bibfnamefont {P.}~\bibnamefont {Jarillo-Herrero}},\ }\href
  {https://doi.org/10.1038/nature26160} {\bibfield  {journal} {\bibinfo
  {journal} {Nature}\ }\textbf {\bibinfo {volume} {556}},\ \bibinfo {pages}
  {43} (\bibinfo {year} {2018}{\natexlab{a}})}\BibitemShut {NoStop}%
\bibitem [{\citenamefont {Cao}\ \emph {et~al.}(2018{\natexlab{b}})\citenamefont
  {Cao}, \citenamefont {Fatemi}, \citenamefont {Demir}, \citenamefont {Fang},
  \citenamefont {Tomarken}, \citenamefont {Luo}, \citenamefont
  {Sanchez-Yamagishi}, \citenamefont {Watanabe}, \citenamefont {Taniguchi},
  \citenamefont {Kaxiras}, \citenamefont {Ashoori},\ and\ \citenamefont
  {Jarillo-Herrero}}]{Cao2018b}%
  \BibitemOpen
  \bibfield  {author} {\bibinfo {author} {\bibfnamefont {Y.}~\bibnamefont
  {Cao}}, \bibinfo {author} {\bibfnamefont {V.}~\bibnamefont {Fatemi}},
  \bibinfo {author} {\bibfnamefont {A.}~\bibnamefont {Demir}}, \bibinfo
  {author} {\bibfnamefont {S.}~\bibnamefont {Fang}}, \bibinfo {author}
  {\bibfnamefont {S.~L.}\ \bibnamefont {Tomarken}}, \bibinfo {author}
  {\bibfnamefont {J.~Y.}\ \bibnamefont {Luo}}, \bibinfo {author} {\bibfnamefont
  {J.~D.}\ \bibnamefont {Sanchez-Yamagishi}}, \bibinfo {author} {\bibfnamefont
  {K.}~\bibnamefont {Watanabe}}, \bibinfo {author} {\bibfnamefont
  {T.}~\bibnamefont {Taniguchi}}, \bibinfo {author} {\bibfnamefont
  {E.}~\bibnamefont {Kaxiras}}, \bibinfo {author} {\bibfnamefont {R.~C.}\
  \bibnamefont {Ashoori}}, \ and\ \bibinfo {author} {\bibfnamefont
  {P.}~\bibnamefont {Jarillo-Herrero}},\ }\href
  {https://doi.org/10.1038/nature26154} {\bibfield  {journal} {\bibinfo
  {journal} {Nature}\ }\textbf {\bibinfo {volume} {556}},\ \bibinfo {pages}
  {80} (\bibinfo {year} {2018}{\natexlab{b}})}\BibitemShut {NoStop}%
\bibitem [{\citenamefont {Yankowitz}\ \emph {et~al.}(2019)\citenamefont
  {Yankowitz}, \citenamefont {Chen}, \citenamefont {Polshyn}, \citenamefont
  {Zhang}, \citenamefont {Watanabe}, \citenamefont {Taniguchi}, \citenamefont
  {Graf}, \citenamefont {Young},\ and\ \citenamefont {Dean}}]{Yankowitz2019}%
  \BibitemOpen
  \bibfield  {author} {\bibinfo {author} {\bibfnamefont {M.}~\bibnamefont
  {Yankowitz}}, \bibinfo {author} {\bibfnamefont {S.}~\bibnamefont {Chen}},
  \bibinfo {author} {\bibfnamefont {H.}~\bibnamefont {Polshyn}}, \bibinfo
  {author} {\bibfnamefont {Y.}~\bibnamefont {Zhang}}, \bibinfo {author}
  {\bibfnamefont {K.}~\bibnamefont {Watanabe}}, \bibinfo {author}
  {\bibfnamefont {T.}~\bibnamefont {Taniguchi}}, \bibinfo {author}
  {\bibfnamefont {D.}~\bibnamefont {Graf}}, \bibinfo {author} {\bibfnamefont
  {A.~F.}\ \bibnamefont {Young}}, \ and\ \bibinfo {author} {\bibfnamefont
  {C.~R.}\ \bibnamefont {Dean}},\ }\href {\doibase 10.1126/science.aav1910}
  {\bibfield  {journal} {\bibinfo  {journal} {Science}\ }\textbf {\bibinfo
  {volume} {363}},\ \bibinfo {pages} {1059} (\bibinfo {year}
  {2019})}\BibitemShut {NoStop}%
\bibitem [{\citenamefont {Huang}\ \emph {et~al.}(2017)\citenamefont {Huang},
  \citenamefont {Clark}, \citenamefont {Navarro-Moratalla}, \citenamefont
  {Klein}, \citenamefont {Cheng}, \citenamefont {Seyler}, \citenamefont
  {Zhong}, \citenamefont {Schmidgall}, \citenamefont {McGuire}, \citenamefont
  {Cobden}, \citenamefont {Yao}, \citenamefont {Xiao}, \citenamefont
  {Jarillo-Herrero},\ and\ \citenamefont {Xu}}]{Huang2017}%
  \BibitemOpen
  \bibfield  {author} {\bibinfo {author} {\bibfnamefont {B.}~\bibnamefont
  {Huang}}, \bibinfo {author} {\bibfnamefont {G.}~\bibnamefont {Clark}},
  \bibinfo {author} {\bibfnamefont {E.}~\bibnamefont {Navarro-Moratalla}},
  \bibinfo {author} {\bibfnamefont {D.~R.}\ \bibnamefont {Klein}}, \bibinfo
  {author} {\bibfnamefont {R.}~\bibnamefont {Cheng}}, \bibinfo {author}
  {\bibfnamefont {K.~L.}\ \bibnamefont {Seyler}}, \bibinfo {author}
  {\bibfnamefont {D.}~\bibnamefont {Zhong}}, \bibinfo {author} {\bibfnamefont
  {E.}~\bibnamefont {Schmidgall}}, \bibinfo {author} {\bibfnamefont {M.~A.}\
  \bibnamefont {McGuire}}, \bibinfo {author} {\bibfnamefont {D.~H.}\
  \bibnamefont {Cobden}}, \bibinfo {author} {\bibfnamefont {W.}~\bibnamefont
  {Yao}}, \bibinfo {author} {\bibfnamefont {D.}~\bibnamefont {Xiao}}, \bibinfo
  {author} {\bibfnamefont {P.}~\bibnamefont {Jarillo-Herrero}}, \ and\ \bibinfo
  {author} {\bibfnamefont {X.}~\bibnamefont {Xu}},\ }\href
  {https://doi.org/10.1038/nature22391} {\bibfield  {journal} {\bibinfo
  {journal} {Nature}\ }\textbf {\bibinfo {volume} {546}},\ \bibinfo {pages}
  {270} (\bibinfo {year} {2017})}\BibitemShut {NoStop}%
\bibitem [{\citenamefont {Bonilla}\ \emph {et~al.}(2018)\citenamefont
  {Bonilla}, \citenamefont {Kolekar}, \citenamefont {Ma}, \citenamefont {Diaz},
  \citenamefont {Kalappattil}, \citenamefont {Das}, \citenamefont {Eggers},
  \citenamefont {Gutierrez}, \citenamefont {Phan},\ and\ \citenamefont
  {Batzill}}]{Bonilla2018}%
  \BibitemOpen
  \bibfield  {author} {\bibinfo {author} {\bibfnamefont {M.}~\bibnamefont
  {Bonilla}}, \bibinfo {author} {\bibfnamefont {S.}~\bibnamefont {Kolekar}},
  \bibinfo {author} {\bibfnamefont {Y.}~\bibnamefont {Ma}}, \bibinfo {author}
  {\bibfnamefont {H.~C.}\ \bibnamefont {Diaz}}, \bibinfo {author}
  {\bibfnamefont {V.}~\bibnamefont {Kalappattil}}, \bibinfo {author}
  {\bibfnamefont {R.}~\bibnamefont {Das}}, \bibinfo {author} {\bibfnamefont
  {T.}~\bibnamefont {Eggers}}, \bibinfo {author} {\bibfnamefont {H.~R.}\
  \bibnamefont {Gutierrez}}, \bibinfo {author} {\bibfnamefont {M.-H.}\
  \bibnamefont {Phan}}, \ and\ \bibinfo {author} {\bibfnamefont
  {M.}~\bibnamefont {Batzill}},\ }\href {\doibase 10.1038/s41565-018-0063-9}
  {\bibfield  {journal} {\bibinfo  {journal} {Nature Nanotech.}\ }\textbf
  {\bibinfo {volume} {13}},\ \bibinfo {pages} {289} (\bibinfo {year}
  {2018})}\BibitemShut {NoStop}%
\bibitem [{\citenamefont {Mak}\ \emph {et~al.}(2012)\citenamefont {Mak},
  \citenamefont {He}, \citenamefont {Lee}, \citenamefont {Lee}, \citenamefont
  {Hone}, \citenamefont {Heinz},\ and\ \citenamefont {Shan}}]{Mak2012}%
  \BibitemOpen
  \bibfield  {author} {\bibinfo {author} {\bibfnamefont {K.~F.}\ \bibnamefont
  {Mak}}, \bibinfo {author} {\bibfnamefont {K.}~\bibnamefont {He}}, \bibinfo
  {author} {\bibfnamefont {C.}~\bibnamefont {Lee}}, \bibinfo {author}
  {\bibfnamefont {G.~H.}\ \bibnamefont {Lee}}, \bibinfo {author} {\bibfnamefont
  {J.}~\bibnamefont {Hone}}, \bibinfo {author} {\bibfnamefont {T.~F.}\
  \bibnamefont {Heinz}}, \ and\ \bibinfo {author} {\bibfnamefont
  {J.}~\bibnamefont {Shan}},\ }\href {https://doi.org/10.1038/nmat3505}
  {\bibfield  {journal} {\bibinfo  {journal} {Nature Mater.}\ }\textbf
  {\bibinfo {volume} {12}},\ \bibinfo {pages} {207} (\bibinfo {year}
  {2012})}\BibitemShut {NoStop}%
\bibitem [{\citenamefont {Britnell}\ \emph {et~al.}(2013)\citenamefont
  {Britnell}, \citenamefont {Ribeiro}, \citenamefont {Eckmann}, \citenamefont
  {Jalil}, \citenamefont {Belle}, \citenamefont {Mishchenko}, \citenamefont
  {Kim}, \citenamefont {Gorbachev}, \citenamefont {Georgiou}, \citenamefont
  {Morozov}, \citenamefont {Grigorenko}, \citenamefont {Geim}, \citenamefont
  {Casiraghi}, \citenamefont {Neto},\ and\ \citenamefont
  {Novoselov}}]{Britnell2013}%
  \BibitemOpen
  \bibfield  {author} {\bibinfo {author} {\bibfnamefont {L.}~\bibnamefont
  {Britnell}}, \bibinfo {author} {\bibfnamefont {R.~M.}\ \bibnamefont
  {Ribeiro}}, \bibinfo {author} {\bibfnamefont {A.}~\bibnamefont {Eckmann}},
  \bibinfo {author} {\bibfnamefont {R.}~\bibnamefont {Jalil}}, \bibinfo
  {author} {\bibfnamefont {B.~D.}\ \bibnamefont {Belle}}, \bibinfo {author}
  {\bibfnamefont {A.}~\bibnamefont {Mishchenko}}, \bibinfo {author}
  {\bibfnamefont {Y.-J.}\ \bibnamefont {Kim}}, \bibinfo {author} {\bibfnamefont
  {R.~V.}\ \bibnamefont {Gorbachev}}, \bibinfo {author} {\bibfnamefont
  {T.}~\bibnamefont {Georgiou}}, \bibinfo {author} {\bibfnamefont {S.~V.}\
  \bibnamefont {Morozov}}, \bibinfo {author} {\bibfnamefont {A.~N.}\
  \bibnamefont {Grigorenko}}, \bibinfo {author} {\bibfnamefont {A.~K.}\
  \bibnamefont {Geim}}, \bibinfo {author} {\bibfnamefont {C.}~\bibnamefont
  {Casiraghi}}, \bibinfo {author} {\bibfnamefont {A.~H.~C.}\ \bibnamefont
  {Neto}}, \ and\ \bibinfo {author} {\bibfnamefont {K.~S.}\ \bibnamefont
  {Novoselov}},\ }\href {\doibase 10.1126/science.1235547} {\bibfield
  {journal} {\bibinfo  {journal} {Science}\ }\textbf {\bibinfo {volume}
  {340}},\ \bibinfo {pages} {1311} (\bibinfo {year} {2013})}\BibitemShut
  {NoStop}%
\bibitem [{\citenamefont {Furchi}\ \emph {et~al.}(2014)\citenamefont {Furchi},
  \citenamefont {Pospischil}, \citenamefont {Libisch}, \citenamefont
  {Burgd{\"o}rfer},\ and\ \citenamefont {Mueller}}]{Furchi2014}%
  \BibitemOpen
  \bibfield  {author} {\bibinfo {author} {\bibfnamefont {M.~M.}\ \bibnamefont
  {Furchi}}, \bibinfo {author} {\bibfnamefont {A.}~\bibnamefont {Pospischil}},
  \bibinfo {author} {\bibfnamefont {F.}~\bibnamefont {Libisch}}, \bibinfo
  {author} {\bibfnamefont {J.}~\bibnamefont {Burgd{\"o}rfer}}, \ and\ \bibinfo
  {author} {\bibfnamefont {T.}~\bibnamefont {Mueller}},\ }\href {\doibase
  10.1021/nl501962c} {\bibfield  {journal} {\bibinfo  {journal} {Nano Lett.}\
  }\textbf {\bibinfo {volume} {14}},\ \bibinfo {pages} {4785} (\bibinfo {year}
  {2014})}\BibitemShut {NoStop}%
\bibitem [{\citenamefont {Sidler}\ \emph {et~al.}(2017)\citenamefont {Sidler},
  \citenamefont {Back}, \citenamefont {Cotlet}, \citenamefont {Srivastava},
  \citenamefont {Fink}, \citenamefont {Kroner}, \citenamefont {Demler},\ and\
  \citenamefont {Imamoglu}}]{Sidler2016}%
  \BibitemOpen
  \bibfield  {author} {\bibinfo {author} {\bibfnamefont {M.}~\bibnamefont
  {Sidler}}, \bibinfo {author} {\bibfnamefont {P.}~\bibnamefont {Back}},
  \bibinfo {author} {\bibfnamefont {O.}~\bibnamefont {Cotlet}}, \bibinfo
  {author} {\bibfnamefont {A.}~\bibnamefont {Srivastava}}, \bibinfo {author}
  {\bibfnamefont {T.}~\bibnamefont {Fink}}, \bibinfo {author} {\bibfnamefont
  {M.}~\bibnamefont {Kroner}}, \bibinfo {author} {\bibfnamefont
  {E.}~\bibnamefont {Demler}}, \ and\ \bibinfo {author} {\bibfnamefont
  {A.}~\bibnamefont {Imamoglu}},\ }\href {\doibase 10.1038/nphys3949}
  {\bibfield  {journal} {\bibinfo  {journal} {Nat. Phys.}\ }\textbf {\bibinfo
  {volume} {13}},\ \bibinfo {pages} {255} (\bibinfo {year} {2017})}\BibitemShut
  {NoStop}%
\bibitem [{\citenamefont {Ravets}\ \emph {et~al.}(2018)\citenamefont {Ravets},
  \citenamefont {Kn\"uppel}, \citenamefont {Faelt}, \citenamefont {Cotlet},
  \citenamefont {Kroner}, \citenamefont {Wegscheider},\ and\ \citenamefont
  {Imamoglu}}]{Ravets2018}%
  \BibitemOpen
  \bibfield  {author} {\bibinfo {author} {\bibfnamefont {S.}~\bibnamefont
  {Ravets}}, \bibinfo {author} {\bibfnamefont {P.}~\bibnamefont {Kn\"uppel}},
  \bibinfo {author} {\bibfnamefont {S.}~\bibnamefont {Faelt}}, \bibinfo
  {author} {\bibfnamefont {O.}~\bibnamefont {Cotlet}}, \bibinfo {author}
  {\bibfnamefont {M.}~\bibnamefont {Kroner}}, \bibinfo {author} {\bibfnamefont
  {W.}~\bibnamefont {Wegscheider}}, \ and\ \bibinfo {author} {\bibfnamefont
  {A.}~\bibnamefont {Imamoglu}},\ }\href {\doibase
  10.1103/PhysRevLett.120.057401} {\bibfield  {journal} {\bibinfo  {journal}
  {Phys. Rev. Lett.}\ }\textbf {\bibinfo {volume} {120}},\ \bibinfo {pages}
  {057401} (\bibinfo {year} {2018})}\BibitemShut {NoStop}%
\bibitem [{\citenamefont {Suris}(2003)}]{Suris2003}%
  \BibitemOpen
  \bibfield  {author} {\bibinfo {author} {\bibfnamefont {R.~A.}\ \bibnamefont
  {Suris}},\ }in\ \href@noop {} {\emph {\bibinfo {booktitle} {Optical
  Properties of 2D Systems with Interacting Electrons}}}\ (\bibinfo
  {publisher} {Springer},\ \bibinfo {year} {2003})\ pp.\ \bibinfo {pages}
  {111--124}\BibitemShut {NoStop}%
\bibitem [{\citenamefont {Prokof'ev}\ and\ \citenamefont
  {Svistunov}(1998{\natexlab{a}})}]{Prokofiev_FrohPolaron1}%
  \BibitemOpen
  \bibfield  {author} {\bibinfo {author} {\bibfnamefont {N.~V.}\ \bibnamefont
  {Prokof'ev}}\ and\ \bibinfo {author} {\bibfnamefont {B.~V.}\ \bibnamefont
  {Svistunov}},\ }\href {\doibase 10.1103/PhysRevLett.81.2514} {\bibfield
  {journal} {\bibinfo  {journal} {Phys. Rev. Lett.}\ }\textbf {\bibinfo
  {volume} {81}},\ \bibinfo {pages} {2514} (\bibinfo {year}
  {1998}{\natexlab{a}})}\BibitemShut {NoStop}%
\bibitem [{\citenamefont {Prokof'ev}\ and\ \citenamefont
  {Svistunov}(1998{\natexlab{b}})}]{Prokofiev_FrohPolaron2}%
  \BibitemOpen
  \bibfield  {author} {\bibinfo {author} {\bibfnamefont {N.~V.}\ \bibnamefont
  {Prokof'ev}}\ and\ \bibinfo {author} {\bibfnamefont {B.~V.}\ \bibnamefont
  {Svistunov}},\ }\href {\doibase 10.1103/PhysRevLett.81.2514} {\bibfield
  {journal} {\bibinfo  {journal} {Phys. Rev. Lett.}\ }\textbf {\bibinfo
  {volume} {81}},\ \bibinfo {pages} {2514} (\bibinfo {year}
  {1998}{\natexlab{b}})}\BibitemShut {NoStop}%
\bibitem [{\citenamefont {Punk}\ \emph {et~al.}(2009)\citenamefont {Punk},
  \citenamefont {Dumitrescu},\ and\ \citenamefont {Zwerger}}]{Punk2009}%
  \BibitemOpen
  \bibfield  {author} {\bibinfo {author} {\bibfnamefont {M.}~\bibnamefont
  {Punk}}, \bibinfo {author} {\bibfnamefont {P.~T.}\ \bibnamefont
  {Dumitrescu}}, \ and\ \bibinfo {author} {\bibfnamefont {W.}~\bibnamefont
  {Zwerger}},\ }\href {\doibase 10.1103/PhysRevA.80.053605} {\bibfield
  {journal} {\bibinfo  {journal} {Phys. Rev. A}\ }\textbf {\bibinfo {volume}
  {80}},\ \bibinfo {pages} {053605} (\bibinfo {year} {2009})}\BibitemShut
  {NoStop}%
\bibitem [{\citenamefont {Schmidt}\ \emph {et~al.}(2012)\citenamefont
  {Schmidt}, \citenamefont {Enss}, \citenamefont {Pietil\"a},\ and\
  \citenamefont {Demler}}]{Schmidt2012b}%
  \BibitemOpen
  \bibfield  {author} {\bibinfo {author} {\bibfnamefont {R.}~\bibnamefont
  {Schmidt}}, \bibinfo {author} {\bibfnamefont {T.}~\bibnamefont {Enss}},
  \bibinfo {author} {\bibfnamefont {V.}~\bibnamefont {Pietil\"a}}, \ and\
  \bibinfo {author} {\bibfnamefont {E.}~\bibnamefont {Demler}},\ }\href
  {\doibase 10.1103/PhysRevA.85.021602} {\bibfield  {journal} {\bibinfo
  {journal} {Phys. Rev. A}\ }\textbf {\bibinfo {volume} {85}},\ \bibinfo
  {pages} {021602} (\bibinfo {year} {2012})}\BibitemShut {NoStop}%
\bibitem [{\citenamefont {Efimkin}\ and\ \citenamefont
  {MacDonald}(2017)}]{Efimkin_Many-body_2017}%
  \BibitemOpen
  \bibfield  {author} {\bibinfo {author} {\bibfnamefont {D.~K.}\ \bibnamefont
  {Efimkin}}\ and\ \bibinfo {author} {\bibfnamefont {A.~H.}\ \bibnamefont
  {MacDonald}},\ }\href {\doibase 10.1103/PhysRevB.95.035417} {\bibfield
  {journal} {\bibinfo  {journal} {Phys. Rev. B}\ }\textbf {\bibinfo {volume}
  {95}},\ \bibinfo {pages} {035417} (\bibinfo {year} {2017})}\BibitemShut
  {NoStop}%
\bibitem [{\citenamefont {Efimkin}\ and\ \citenamefont
  {MacDonald}(2018)}]{Efimkin_Exciton-polarons_2018}%
  \BibitemOpen
  \bibfield  {author} {\bibinfo {author} {\bibfnamefont {D.~K.}\ \bibnamefont
  {Efimkin}}\ and\ \bibinfo {author} {\bibfnamefont {A.~H.}\ \bibnamefont
  {MacDonald}},\ }\href {\doibase 10.1103/PhysRevB.97.235432} {\bibfield
  {journal} {\bibinfo  {journal} {Phys. Rev. B}\ }\textbf {\bibinfo {volume}
  {97}},\ \bibinfo {pages} {235432} (\bibinfo {year} {2018})}\BibitemShut
  {NoStop}%
\bibitem [{\citenamefont {Cotle}\ \emph {et~al.}(2019)\citenamefont {Cotle},
  \citenamefont {Pientka}, \citenamefont {Schmidt}, \citenamefont {Zarand},
  \citenamefont {Demler},\ and\ \citenamefont {Imamoglu}}]{cotlet2018}%
  \BibitemOpen
  \bibfield  {author} {\bibinfo {author} {\bibfnamefont {O.}~\bibnamefont
  {Cotle}}, \bibinfo {author} {\bibfnamefont {F.}~\bibnamefont {Pientka}},
  \bibinfo {author} {\bibfnamefont {R.}~\bibnamefont {Schmidt}}, \bibinfo
  {author} {\bibfnamefont {G.}~\bibnamefont {Zarand}}, \bibinfo {author}
  {\bibfnamefont {E.}~\bibnamefont {Demler}}, \ and\ \bibinfo {author}
  {\bibfnamefont {A.}~\bibnamefont {Imamoglu}},\ }\href {\doibase
  10.1103/PhysRevX.9.041019} {\bibfield  {journal} {\bibinfo  {journal} {Phys.
  Rev. X}\ }\textbf {\bibinfo {volume} {9}},\ \bibinfo {pages} {041019}
  (\bibinfo {year} {2019})}\BibitemShut {NoStop}%
\bibitem [{\citenamefont {Schmidt}\ \emph {et~al.}(2018)\citenamefont
  {Schmidt}, \citenamefont {Knap}, \citenamefont {Ivanov}, \citenamefont {You},
  \citenamefont {Cetina},\ and\ \citenamefont {Demler}}]{schmidt2017}%
  \BibitemOpen
  \bibfield  {author} {\bibinfo {author} {\bibfnamefont {R.}~\bibnamefont
  {Schmidt}}, \bibinfo {author} {\bibfnamefont {M.}~\bibnamefont {Knap}},
  \bibinfo {author} {\bibfnamefont {D.~A.}\ \bibnamefont {Ivanov}}, \bibinfo
  {author} {\bibfnamefont {J.-S.}\ \bibnamefont {You}}, \bibinfo {author}
  {\bibfnamefont {M.}~\bibnamefont {Cetina}}, \ and\ \bibinfo {author}
  {\bibfnamefont {E.}~\bibnamefont {Demler}},\ }\href
  {http://stacks.iop.org/0034-4885/81/i=2/a=024401} {\bibfield  {journal}
  {\bibinfo  {journal} {Rep. Prog. Phys.}\ }\textbf {\bibinfo {volume} {81}},\
  \bibinfo {pages} {024401} (\bibinfo {year} {2018})}\BibitemShut {NoStop}%
\bibitem [{\citenamefont {Anderson}(1967)}]{Anderson1967}%
  \BibitemOpen
  \bibfield  {author} {\bibinfo {author} {\bibfnamefont {P.~W.}\ \bibnamefont
  {Anderson}},\ }\href {\doibase 10.1103/physrevlett.18.1049} {\bibfield
  {journal} {\bibinfo  {journal} {Phys. Rev. Lett.}\ }\textbf {\bibinfo
  {volume} {18}},\ \bibinfo {pages} {1049} (\bibinfo {year}
  {1967})}\BibitemShut {NoStop}%
\bibitem [{\citenamefont {Rosch}\ and\ \citenamefont {Kopp}(1995)}]{AchimKopp}%
  \BibitemOpen
  \bibfield  {author} {\bibinfo {author} {\bibfnamefont {A.}~\bibnamefont
  {Rosch}}\ and\ \bibinfo {author} {\bibfnamefont {T.}~\bibnamefont {Kopp}},\
  }\href {\doibase 10.1103/PhysRevLett.75.1988} {\bibfield  {journal} {\bibinfo
   {journal} {Phys. Rev. Lett.}\ }\textbf {\bibinfo {volume} {75}},\ \bibinfo
  {pages} {1988} (\bibinfo {year} {1995})}\BibitemShut {NoStop}%
\bibitem [{\citenamefont {Kondo}\ and\ \citenamefont {Soda}(1983)}]{Kondo1983}%
  \BibitemOpen
  \bibfield  {author} {\bibinfo {author} {\bibfnamefont {J.}~\bibnamefont
  {Kondo}}\ and\ \bibinfo {author} {\bibfnamefont {T.}~\bibnamefont {Soda}},\
  }\href@noop {} {\bibfield  {journal} {\bibinfo  {journal} {Journal of Low
  Temperature Physics}\ }\textbf {\bibinfo {volume} {50}},\ \bibinfo {pages}
  {21} (\bibinfo {year} {1983})}\BibitemShut {NoStop}%
\bibitem [{\citenamefont {Tan}\ \emph {et~al.}(2019)\citenamefont {Tan},
  \citenamefont {Cotlet}, \citenamefont {Bergschneider}, \citenamefont
  {Schmidt}, \citenamefont {Back}, \citenamefont {Shimazaki}, \citenamefont
  {Kroner},\ and\ \citenamefont {Imamoglu}}]{Tan2019}%
  \BibitemOpen
  \bibfield  {author} {\bibinfo {author} {\bibfnamefont {L.~B.}\ \bibnamefont
  {Tan}}, \bibinfo {author} {\bibfnamefont {O.}~\bibnamefont {Cotlet}},
  \bibinfo {author} {\bibfnamefont {A.}~\bibnamefont {Bergschneider}}, \bibinfo
  {author} {\bibfnamefont {R.}~\bibnamefont {Schmidt}}, \bibinfo {author}
  {\bibfnamefont {P.}~\bibnamefont {Back}}, \bibinfo {author} {\bibfnamefont
  {Y.}~\bibnamefont {Shimazaki}}, \bibinfo {author} {\bibfnamefont
  {M.}~\bibnamefont {Kroner}}, \ and\ \bibinfo {author} {\bibfnamefont
  {A.}~\bibnamefont {Imamoglu}},\ }\href@noop {} {\bibfield  {journal}
  {\bibinfo  {journal} {arXiv:1903.05640}\ } (\bibinfo {year}
  {2019})}\BibitemShut {NoStop}%
\bibitem [{\citenamefont {Cotle\ifmmode~\mbox{\c{t}}\else \c{t}\fi{}}\ \emph
  {et~al.}(2016)\citenamefont {Cotle\ifmmode~\mbox{\c{t}}\else \c{t}\fi{}},
  \citenamefont {Zeytino\ifmmode~\check{g}\else \v{g}\fi{}lu}, \citenamefont
  {Sigrist}, \citenamefont {Demler},\ and\ \citenamefont
  {Imamo\ifmmode~\check{g}\else \v{g}\fi{}lu}}]{Cotlet2016}%
  \BibitemOpen
  \bibfield  {author} {\bibinfo {author} {\bibfnamefont {O.}~\bibnamefont
  {Cotle\ifmmode~\mbox{\c{t}}\else \c{t}\fi{}}}, \bibinfo {author}
  {\bibfnamefont {S.}~\bibnamefont {Zeytino\ifmmode~\check{g}\else
  \v{g}\fi{}lu}}, \bibinfo {author} {\bibfnamefont {M.}~\bibnamefont
  {Sigrist}}, \bibinfo {author} {\bibfnamefont {E.}~\bibnamefont {Demler}}, \
  and\ \bibinfo {author} {\bibfnamefont {A.~m.~c.}\ \bibnamefont
  {Imamo\ifmmode~\check{g}\else \v{g}\fi{}lu}},\ }\href {\doibase
  10.1103/PhysRevB.93.054510} {\bibfield  {journal} {\bibinfo  {journal} {Phys.
  Rev. B}\ }\textbf {\bibinfo {volume} {93}},\ \bibinfo {pages} {054510}
  (\bibinfo {year} {2016})}\BibitemShut {NoStop}%
\bibitem [{\citenamefont {Kavokin}\ and\ \citenamefont
  {Lagoudakis}(2016)}]{Kavokin2016}%
  \BibitemOpen
  \bibfield  {author} {\bibinfo {author} {\bibfnamefont {A.}~\bibnamefont
  {Kavokin}}\ and\ \bibinfo {author} {\bibfnamefont {P.}~\bibnamefont
  {Lagoudakis}},\ }\href {https://doi.org/10.1038/nmat4646} {\bibfield
  {journal} {\bibinfo  {journal} {Nature Mater.}\ }\textbf {\bibinfo {volume}
  {15}},\ \bibinfo {pages} {599} (\bibinfo {year} {2016})}\BibitemShut
  {NoStop}%
\bibitem [{\citenamefont {Shelykh}\ \emph {et~al.}(2010)\citenamefont
  {Shelykh}, \citenamefont {Taylor},\ and\ \citenamefont
  {Kavokin}}]{Shelykh2010}%
  \BibitemOpen
  \bibfield  {author} {\bibinfo {author} {\bibfnamefont {I.~A.}\ \bibnamefont
  {Shelykh}}, \bibinfo {author} {\bibfnamefont {T.}~\bibnamefont {Taylor}}, \
  and\ \bibinfo {author} {\bibfnamefont {A.~V.}\ \bibnamefont {Kavokin}},\
  }\href {\doibase 10.1103/PhysRevLett.105.140402} {\bibfield  {journal}
  {\bibinfo  {journal} {Phys. Rev. Lett.}\ }\textbf {\bibinfo {volume} {105}},\
  \bibinfo {pages} {140402} (\bibinfo {year} {2010})}\BibitemShut {NoStop}%
\bibitem [{\citenamefont {Berkelbach}\ \emph {et~al.}(2013)\citenamefont
  {Berkelbach}, \citenamefont {Hybertsen},\ and\ \citenamefont
  {Reichman}}]{berkelbach_theory_2013}%
  \BibitemOpen
  \bibfield  {author} {\bibinfo {author} {\bibfnamefont {T.~C.}\ \bibnamefont
  {Berkelbach}}, \bibinfo {author} {\bibfnamefont {M.~S.}\ \bibnamefont
  {Hybertsen}}, \ and\ \bibinfo {author} {\bibfnamefont {D.~R.}\ \bibnamefont
  {Reichman}},\ }\href {\doibase 10.1103/PhysRevB.88.045318} {\bibfield
  {journal} {\bibinfo  {journal} {Phys. Rev. B}\ }\textbf {\bibinfo {volume}
  {88}},\ \bibinfo {pages} {045318} (\bibinfo {year} {2013})}\BibitemShut
  {NoStop}%
\bibitem [{\citenamefont {Courtade}\ \emph {et~al.}(2017)\citenamefont
  {Courtade}, \citenamefont {Semina}, \citenamefont {Manca}, \citenamefont
  {Glazov}, \citenamefont {Robert}, \citenamefont {Cadiz}, \citenamefont
  {Wang}, \citenamefont {Taniguchi}, \citenamefont {Watanabe}, \citenamefont
  {Pierre}, \citenamefont {Escoffier}, \citenamefont {Ivchenko}, \citenamefont
  {Renucci}, \citenamefont {Marie}, \citenamefont {Amand},\ and\ \citenamefont
  {Urbaszek}}]{Courtade_Chraged_excitons_2017}%
  \BibitemOpen
  \bibfield  {author} {\bibinfo {author} {\bibfnamefont {E.}~\bibnamefont
  {Courtade}}, \bibinfo {author} {\bibfnamefont {M.}~\bibnamefont {Semina}},
  \bibinfo {author} {\bibfnamefont {M.}~\bibnamefont {Manca}}, \bibinfo
  {author} {\bibfnamefont {M.~M.}\ \bibnamefont {Glazov}}, \bibinfo {author}
  {\bibfnamefont {C.}~\bibnamefont {Robert}}, \bibinfo {author} {\bibfnamefont
  {F.}~\bibnamefont {Cadiz}}, \bibinfo {author} {\bibfnamefont
  {G.}~\bibnamefont {Wang}}, \bibinfo {author} {\bibfnamefont {T.}~\bibnamefont
  {Taniguchi}}, \bibinfo {author} {\bibfnamefont {K.}~\bibnamefont {Watanabe}},
  \bibinfo {author} {\bibfnamefont {M.}~\bibnamefont {Pierre}}, \bibinfo
  {author} {\bibfnamefont {W.}~\bibnamefont {Escoffier}}, \bibinfo {author}
  {\bibfnamefont {E.~L.}\ \bibnamefont {Ivchenko}}, \bibinfo {author}
  {\bibfnamefont {P.}~\bibnamefont {Renucci}}, \bibinfo {author} {\bibfnamefont
  {X.}~\bibnamefont {Marie}}, \bibinfo {author} {\bibfnamefont
  {T.}~\bibnamefont {Amand}}, \ and\ \bibinfo {author} {\bibfnamefont
  {B.}~\bibnamefont {Urbaszek}},\ }\href {\doibase 10.1103/PhysRevB.96.085302}
  {\bibfield  {journal} {\bibinfo  {journal} {Phys. Rev. B}\ }\textbf {\bibinfo
  {volume} {96}},\ \bibinfo {pages} {085302} (\bibinfo {year}
  {2017})}\BibitemShut {NoStop}%
\bibitem [{\citenamefont {Keldysh}(1979)}]{keldysh1979}%
  \BibitemOpen
  \bibfield  {author} {\bibinfo {author} {\bibfnamefont {L.~V.}\ \bibnamefont
  {Keldysh}},\ }\href@noop {} {\bibfield  {journal} {\bibinfo  {journal} {JETP
  Lett.}\ }\textbf {\bibinfo {volume} {29}},\ \bibinfo {pages} {658} (\bibinfo
  {year} {1979})}\BibitemShut {NoStop}%
\bibitem [{\citenamefont {Cudazzo}\ \emph {et~al.}(2011)\citenamefont
  {Cudazzo}, \citenamefont {Tokatly},\ and\ \citenamefont
  {Rubio}}]{Cudazzo_dielectric_screening_2011}%
  \BibitemOpen
  \bibfield  {author} {\bibinfo {author} {\bibfnamefont {P.}~\bibnamefont
  {Cudazzo}}, \bibinfo {author} {\bibfnamefont {I.~V.}\ \bibnamefont
  {Tokatly}}, \ and\ \bibinfo {author} {\bibfnamefont {A.}~\bibnamefont
  {Rubio}},\ }\href {\doibase 10.1103/PhysRevB.84.085406} {\bibfield  {journal}
  {\bibinfo  {journal} {Phys. Rev. B}\ }\textbf {\bibinfo {volume} {84}},\
  \bibinfo {pages} {085406} (\bibinfo {year} {2011})}\BibitemShut {NoStop}%
\bibitem [{\citenamefont {Kyl{\"a}np{\"a}{\"a}}\ and\ \citenamefont
  {Komsa}(2015)}]{kylanpaa_binding_2015}%
  \BibitemOpen
  \bibfield  {author} {\bibinfo {author} {\bibfnamefont {I.}~\bibnamefont
  {Kyl{\"a}np{\"a}{\"a}}}\ and\ \bibinfo {author} {\bibfnamefont {H.-P.}\
  \bibnamefont {Komsa}},\ }\href {\doibase 10.1103/PhysRevB.92.205418}
  {\bibfield  {journal} {\bibinfo  {journal} {Phys. Rev. B}\ }\textbf {\bibinfo
  {volume} {92}},\ \bibinfo {pages} {205418} (\bibinfo {year}
  {2015})}\BibitemShut {NoStop}%
\bibitem [{\citenamefont {McCurdy}\ \emph {et~al.}(2003)\citenamefont
  {McCurdy}, \citenamefont {Isaacs}, \citenamefont {Meyer},\ and\ \citenamefont
  {Rescigno}}]{mccurdy_resonant_2003}%
  \BibitemOpen
  \bibfield  {author} {\bibinfo {author} {\bibfnamefont {C.~W.}\ \bibnamefont
  {McCurdy}}, \bibinfo {author} {\bibfnamefont {W.~A.}\ \bibnamefont {Isaacs}},
  \bibinfo {author} {\bibfnamefont {H.-D.}\ \bibnamefont {Meyer}}, \ and\
  \bibinfo {author} {\bibfnamefont {T.~N.}\ \bibnamefont {Rescigno}},\ }\href
  {https://doi.org/10.1103/PhysRevA.67.042708} {\bibfield  {journal} {\bibinfo
  {journal} {Phys. Rev. A}\ }\textbf {\bibinfo {volume} {67}} (\bibinfo {year}
  {2003})}\BibitemShut {NoStop}%
\bibitem [{\citenamefont {Beck}\ and\ \citenamefont
  {Meyer}(2000)}]{beck_multiconfiguration_2000}%
  \BibitemOpen
  \bibfield  {author} {\bibinfo {author} {\bibfnamefont {M.}~\bibnamefont
  {Beck}}\ and\ \bibinfo {author} {\bibfnamefont {H.-D.}\ \bibnamefont
  {Meyer}},\ }\href {\doibase 10.1016/S0370-1573(99)00047-2} {\bibfield
  {journal} {\bibinfo  {journal} {Phys. Rep.}\ }\textbf {\bibinfo {volume}
  {324}},\ \bibinfo {pages} {1} (\bibinfo {year} {2000})}\BibitemShut {NoStop}%
\bibitem [{\citenamefont {Srivastava}\ and\ \citenamefont
  {Imamoglu}(2015)}]{srivastava2015}%
  \BibitemOpen
  \bibfield  {author} {\bibinfo {author} {\bibfnamefont {A.}~\bibnamefont
  {Srivastava}}\ and\ \bibinfo {author} {\bibfnamefont {A.}~\bibnamefont
  {Imamoglu}},\ }\href {\doibase 10.1103/PhysRevLett.115.166802} {\bibfield
  {journal} {\bibinfo  {journal} {Phys. Rev. Lett.}\ }\textbf {\bibinfo
  {volume} {115}},\ \bibinfo {pages} {166802} (\bibinfo {year}
  {2015})}\BibitemShut {NoStop}%
\bibitem [{\citenamefont {Glazov}\ \emph {et~al.}(2015)\citenamefont {Glazov},
  \citenamefont {Ivchenko}, \citenamefont {Wang}, \citenamefont {Amand},
  \citenamefont {Marie}, \citenamefont {Urbaszek},\ and\ \citenamefont
  {Liu}}]{Glazov_2015}%
  \BibitemOpen
  \bibfield  {author} {\bibinfo {author} {\bibfnamefont {M.~M.}\ \bibnamefont
  {Glazov}}, \bibinfo {author} {\bibfnamefont {E.~L.}\ \bibnamefont
  {Ivchenko}}, \bibinfo {author} {\bibfnamefont {G.}~\bibnamefont {Wang}},
  \bibinfo {author} {\bibfnamefont {T.}~\bibnamefont {Amand}}, \bibinfo
  {author} {\bibfnamefont {X.}~\bibnamefont {Marie}}, \bibinfo {author}
  {\bibfnamefont {B.}~\bibnamefont {Urbaszek}}, \ and\ \bibinfo {author}
  {\bibfnamefont {B.~L.}\ \bibnamefont {Liu}},\ }\href {\doibase
  10.1002/pssb.201552211} {\bibfield  {journal} {\bibinfo  {journal} {Phys.
  Status Solidi B}\ }\textbf {\bibinfo {volume} {252}},\ \bibinfo {pages}
  {2349–2362} (\bibinfo {year} {2015})}\BibitemShut {NoStop}%
\bibitem [{\citenamefont {Plechinger}\ \emph {et~al.}(2016)\citenamefont
  {Plechinger}, \citenamefont {Nagler}, \citenamefont {Arora}, \citenamefont
  {Schmidt}, \citenamefont {Chernikov}, \citenamefont {del Águila},
  \citenamefont {Christianen}, \citenamefont {Bratschitsch}, \citenamefont
  {Schüller},\ and\ \citenamefont {Korn}}]{Plechinger_2016}%
  \BibitemOpen
  \bibfield  {author} {\bibinfo {author} {\bibfnamefont {G.}~\bibnamefont
  {Plechinger}}, \bibinfo {author} {\bibfnamefont {P.}~\bibnamefont {Nagler}},
  \bibinfo {author} {\bibfnamefont {A.}~\bibnamefont {Arora}}, \bibinfo
  {author} {\bibfnamefont {R.}~\bibnamefont {Schmidt}}, \bibinfo {author}
  {\bibfnamefont {A.}~\bibnamefont {Chernikov}}, \bibinfo {author}
  {\bibfnamefont {A.~G.}\ \bibnamefont {del Águila}}, \bibinfo {author}
  {\bibfnamefont {P.~C.}\ \bibnamefont {Christianen}}, \bibinfo {author}
  {\bibfnamefont {R.}~\bibnamefont {Bratschitsch}}, \bibinfo {author}
  {\bibfnamefont {C.}~\bibnamefont {Schüller}}, \ and\ \bibinfo {author}
  {\bibfnamefont {T.}~\bibnamefont {Korn}},\ }\href {\doibase
  10.1038/ncomms12715} {\bibfield  {journal} {\bibinfo  {journal} {Nature
  Commun.}\ }\textbf {\bibinfo {volume} {7}},\ \bibinfo {pages} {12715}
  (\bibinfo {year} {2016})}\BibitemShut {NoStop}%
\bibitem [{\citenamefont {Sergeev}\ and\ \citenamefont
  {Suris}(2001)}]{sergeev2001triplet}%
  \BibitemOpen
  \bibfield  {author} {\bibinfo {author} {\bibfnamefont {R.}~\bibnamefont
  {Sergeev}}\ and\ \bibinfo {author} {\bibfnamefont {R.}~\bibnamefont
  {Suris}},\ }\href@noop {} {\bibfield  {journal} {\bibinfo  {journal} {Phys.
  Status Solidi B}\ }\textbf {\bibinfo {volume} {227}},\ \bibinfo {pages} {387}
  (\bibinfo {year} {2001})}\BibitemShut {NoStop}%
\bibitem [{\citenamefont {Wang}\ \emph {et~al.}(2018)\citenamefont {Wang},
  \citenamefont {Chernikov}, \citenamefont {Glazov}, \citenamefont {Heinz},
  \citenamefont {Marie}, \citenamefont {Amand},\ and\ \citenamefont
  {Urbaszek}}]{wang2018}%
  \BibitemOpen
  \bibfield  {author} {\bibinfo {author} {\bibfnamefont {G.}~\bibnamefont
  {Wang}}, \bibinfo {author} {\bibfnamefont {A.}~\bibnamefont {Chernikov}},
  \bibinfo {author} {\bibfnamefont {M.~M.}\ \bibnamefont {Glazov}}, \bibinfo
  {author} {\bibfnamefont {T.~F.}\ \bibnamefont {Heinz}}, \bibinfo {author}
  {\bibfnamefont {X.}~\bibnamefont {Marie}}, \bibinfo {author} {\bibfnamefont
  {T.}~\bibnamefont {Amand}}, \ and\ \bibinfo {author} {\bibfnamefont
  {B.}~\bibnamefont {Urbaszek}},\ }\href {\doibase
  10.1103/RevModPhys.90.021001} {\bibfield  {journal} {\bibinfo  {journal}
  {Rev. Mod. Phys.}\ }\textbf {\bibinfo {volume} {90}},\ \bibinfo {pages}
  {021001} (\bibinfo {year} {2018})}\BibitemShut {NoStop}%
\bibitem [{\citenamefont {Chernikov}\ \emph {et~al.}(2014)\citenamefont
  {Chernikov}, \citenamefont {Berkelbach}, \citenamefont {Hill}, \citenamefont
  {Rigosi}, \citenamefont {Li}, \citenamefont {Aslan}, \citenamefont
  {Reichman}, \citenamefont {Hybertsen},\ and\ \citenamefont
  {Heinz}}]{chernikov2014}%
  \BibitemOpen
  \bibfield  {author} {\bibinfo {author} {\bibfnamefont {A.}~\bibnamefont
  {Chernikov}}, \bibinfo {author} {\bibfnamefont {T.~C.}\ \bibnamefont
  {Berkelbach}}, \bibinfo {author} {\bibfnamefont {H.~M.}\ \bibnamefont
  {Hill}}, \bibinfo {author} {\bibfnamefont {A.}~\bibnamefont {Rigosi}},
  \bibinfo {author} {\bibfnamefont {Y.}~\bibnamefont {Li}}, \bibinfo {author}
  {\bibfnamefont {O.~B.}\ \bibnamefont {Aslan}}, \bibinfo {author}
  {\bibfnamefont {D.~R.}\ \bibnamefont {Reichman}}, \bibinfo {author}
  {\bibfnamefont {M.~S.}\ \bibnamefont {Hybertsen}}, \ and\ \bibinfo {author}
  {\bibfnamefont {T.~F.}\ \bibnamefont {Heinz}},\ }\href {\doibase
  10.1103/PhysRevLett.113.076802} {\bibfield  {journal} {\bibinfo  {journal}
  {Phys. Rev. Lett.}\ }\textbf {\bibinfo {volume} {113}},\ \bibinfo {pages}
  {076802} (\bibinfo {year} {2014})}\BibitemShut {NoStop}%
\bibitem [{\citenamefont {Robert}\ \emph {et~al.}(2018)\citenamefont {Robert},
  \citenamefont {Semina}, \citenamefont {Cadiz}, \citenamefont {Manca},
  \citenamefont {Courtade}, \citenamefont {Taniguchi}, \citenamefont
  {Watanabe}, \citenamefont {Cai}, \citenamefont {Tongay}, \citenamefont
  {Lassagne}, \citenamefont {Renucci}, \citenamefont {Amand}, \citenamefont
  {Marie}, \citenamefont {Glazov},\ and\ \citenamefont
  {Urbaszek}}]{robert2018}%
  \BibitemOpen
  \bibfield  {author} {\bibinfo {author} {\bibfnamefont {C.}~\bibnamefont
  {Robert}}, \bibinfo {author} {\bibfnamefont {M.~A.}\ \bibnamefont {Semina}},
  \bibinfo {author} {\bibfnamefont {F.}~\bibnamefont {Cadiz}}, \bibinfo
  {author} {\bibfnamefont {M.}~\bibnamefont {Manca}}, \bibinfo {author}
  {\bibfnamefont {E.}~\bibnamefont {Courtade}}, \bibinfo {author}
  {\bibfnamefont {T.}~\bibnamefont {Taniguchi}}, \bibinfo {author}
  {\bibfnamefont {K.}~\bibnamefont {Watanabe}}, \bibinfo {author}
  {\bibfnamefont {H.}~\bibnamefont {Cai}}, \bibinfo {author} {\bibfnamefont
  {S.}~\bibnamefont {Tongay}}, \bibinfo {author} {\bibfnamefont
  {B.}~\bibnamefont {Lassagne}}, \bibinfo {author} {\bibfnamefont
  {P.}~\bibnamefont {Renucci}}, \bibinfo {author} {\bibfnamefont
  {T.}~\bibnamefont {Amand}}, \bibinfo {author} {\bibfnamefont
  {X.}~\bibnamefont {Marie}}, \bibinfo {author} {\bibfnamefont {M.~M.}\
  \bibnamefont {Glazov}}, \ and\ \bibinfo {author} {\bibfnamefont
  {B.}~\bibnamefont {Urbaszek}},\ }\href {\doibase
  10.1103/PhysRevMaterials.2.011001} {\bibfield  {journal} {\bibinfo  {journal}
  {Phys. Rev. Materials}\ }\textbf {\bibinfo {volume} {2}},\ \bibinfo {pages}
  {011001} (\bibinfo {year} {2018})}\BibitemShut {NoStop}%
\bibitem [{\citenamefont {Florian}\ \emph {et~al.}(2018)\citenamefont
  {Florian}, \citenamefont {Hartmann}, \citenamefont {Steinhoff}, \citenamefont
  {Klein}, \citenamefont {Holleitner}, \citenamefont {Finley}, \citenamefont
  {Wehling}, \citenamefont {Kaniber},\ and\ \citenamefont
  {Gies}}]{florian2018}%
  \BibitemOpen
  \bibfield  {author} {\bibinfo {author} {\bibfnamefont {M.}~\bibnamefont
  {Florian}}, \bibinfo {author} {\bibfnamefont {M.}~\bibnamefont {Hartmann}},
  \bibinfo {author} {\bibfnamefont {A.}~\bibnamefont {Steinhoff}}, \bibinfo
  {author} {\bibfnamefont {J.}~\bibnamefont {Klein}}, \bibinfo {author}
  {\bibfnamefont {A.~W.}\ \bibnamefont {Holleitner}}, \bibinfo {author}
  {\bibfnamefont {J.~J.}\ \bibnamefont {Finley}}, \bibinfo {author}
  {\bibfnamefont {T.~O.}\ \bibnamefont {Wehling}}, \bibinfo {author}
  {\bibfnamefont {M.}~\bibnamefont {Kaniber}}, \ and\ \bibinfo {author}
  {\bibfnamefont {C.}~\bibnamefont {Gies}},\ }\href {\doibase
  10.1021/acs.nanolett.8b00840} {\bibfield  {journal} {\bibinfo  {journal}
  {Nano Lett.}\ }\textbf {\bibinfo {volume} {18}},\ \bibinfo {pages} {2725}
  (\bibinfo {year} {2018})}\BibitemShut {NoStop}%
\bibitem [{\citenamefont {Arora}\ \emph {et~al.}(2019)\citenamefont {Arora},
  \citenamefont {Deilmann}, \citenamefont {Reichenauer}, \citenamefont {Kern},
  \citenamefont {Michaelis~de Vasconcellos}, \citenamefont {Rohlfing},\ and\
  \citenamefont {Bratschitsch}}]{Arora_2019}%
  \BibitemOpen
  \bibfield  {author} {\bibinfo {author} {\bibfnamefont {A.}~\bibnamefont
  {Arora}}, \bibinfo {author} {\bibfnamefont {T.}~\bibnamefont {Deilmann}},
  \bibinfo {author} {\bibfnamefont {T.}~\bibnamefont {Reichenauer}}, \bibinfo
  {author} {\bibfnamefont {J.}~\bibnamefont {Kern}}, \bibinfo {author}
  {\bibfnamefont {S.}~\bibnamefont {Michaelis~de Vasconcellos}}, \bibinfo
  {author} {\bibfnamefont {M.}~\bibnamefont {Rohlfing}}, \ and\ \bibinfo
  {author} {\bibfnamefont {R.}~\bibnamefont {Bratschitsch}},\ }\href {\doibase
  10.1103/PhysRevLett.123.167401} {\bibfield  {journal} {\bibinfo  {journal}
  {Phys. Rev. Lett.}\ }\textbf {\bibinfo {volume} {123}},\ \bibinfo {pages}
  {167401} (\bibinfo {year} {2019})}\BibitemShut {NoStop}%
\bibitem [{\citenamefont {Shiau}\ \emph {et~al.}(2012)\citenamefont {Shiau},
  \citenamefont {Combescot},\ and\ \citenamefont
  {Chang}}]{Shiau_Combescot_Chang_2012}%
  \BibitemOpen
  \bibfield  {author} {\bibinfo {author} {\bibfnamefont {S.-Y.}\ \bibnamefont
  {Shiau}}, \bibinfo {author} {\bibfnamefont {M.}~\bibnamefont {Combescot}}, \
  and\ \bibinfo {author} {\bibfnamefont {Y.-C.}\ \bibnamefont {Chang}},\ }\href
  {\doibase 10.1103/PhysRevB.86.115210} {\bibfield  {journal} {\bibinfo
  {journal} {Phys. Rev. B}\ }\textbf {\bibinfo {volume} {86}} (\bibinfo {year}
  {2012}),\ 10.1103/PhysRevB.86.115210}\BibitemShut {NoStop}%
\bibitem [{\citenamefont {Back}\ \emph {et~al.}(2017)\citenamefont {Back},
  \citenamefont {Sidler}, \citenamefont {Cotlet}, \citenamefont {Srivastava},
  \citenamefont {Takemura}, \citenamefont {Kroner},\ and\ \citenamefont
  {Imamoglu}}]{Back2017}%
  \BibitemOpen
  \bibfield  {author} {\bibinfo {author} {\bibfnamefont {P.}~\bibnamefont
  {Back}}, \bibinfo {author} {\bibfnamefont {M.}~\bibnamefont {Sidler}},
  \bibinfo {author} {\bibfnamefont {O.}~\bibnamefont {Cotlet}}, \bibinfo
  {author} {\bibfnamefont {A.}~\bibnamefont {Srivastava}}, \bibinfo {author}
  {\bibfnamefont {N.}~\bibnamefont {Takemura}}, \bibinfo {author}
  {\bibfnamefont {M.}~\bibnamefont {Kroner}}, \ and\ \bibinfo {author}
  {\bibfnamefont {A.}~\bibnamefont {Imamoglu}},\ }\href {\doibase
  10.1103/PhysRevLett.118.237404} {\bibfield  {journal} {\bibinfo  {journal}
  {Phys. Rev. Lett.}\ }\textbf {\bibinfo {volume} {118}},\ \bibinfo {pages}
  {237404} (\bibinfo {year} {2017})}\BibitemShut {NoStop}%
\bibitem [{\citenamefont {Verhaar}\ \emph {et~al.}(1984)\citenamefont
  {Verhaar}, \citenamefont {van~den Eijnde}, \citenamefont {Voermans},\ and\
  \citenamefont {Schaffrath}}]{verhaar_scattering_1984}%
  \BibitemOpen
  \bibfield  {author} {\bibinfo {author} {\bibfnamefont {B.~J.}\ \bibnamefont
  {Verhaar}}, \bibinfo {author} {\bibfnamefont {J.}~\bibnamefont {van~den
  Eijnde}}, \bibinfo {author} {\bibfnamefont {M.~A.~J.}\ \bibnamefont
  {Voermans}}, \ and\ \bibinfo {author} {\bibfnamefont {M.~M.~J.}\ \bibnamefont
  {Schaffrath}},\ }\href {\doibase 10.1088/0305-4470/17/3/020} {\bibfield
  {journal} {\bibinfo  {journal} {J. Phys. A}\ }\textbf {\bibinfo {volume}
  {17}},\ \bibinfo {pages} {595} (\bibinfo {year} {1984})}\BibitemShut
  {NoStop}%
\bibitem [{\citenamefont {Adhikari}(1986)}]{adhikari_quantum_1986}%
  \BibitemOpen
  \bibfield  {author} {\bibinfo {author} {\bibfnamefont {S.~K.}\ \bibnamefont
  {Adhikari}},\ }\href {\doibase 10.1119/1.14623} {\bibfield  {journal}
  {\bibinfo  {journal} {Am. J. Phys.}\ }\textbf {\bibinfo {volume} {54}},\
  \bibinfo {pages} {362} (\bibinfo {year} {1986})}\BibitemShut {NoStop}%
\bibitem [{\citenamefont {Astrakharchik}\ \emph {et~al.}(2004)\citenamefont
  {Astrakharchik}, \citenamefont {Boronat}, \citenamefont {Casulleras},\ and\
  \citenamefont {Giorgini}}]{Astrakharchik2004}%
  \BibitemOpen
  \bibfield  {author} {\bibinfo {author} {\bibfnamefont {G.~E.}\ \bibnamefont
  {Astrakharchik}}, \bibinfo {author} {\bibfnamefont {J.}~\bibnamefont
  {Boronat}}, \bibinfo {author} {\bibfnamefont {J.}~\bibnamefont {Casulleras}},
  \ and\ \bibinfo {author} {\bibfnamefont {S.}~\bibnamefont {Giorgini}},\
  }\href {\doibase 10.1103/PhysRevLett.93.200404} {\bibfield  {journal}
  {\bibinfo  {journal} {Phys. Rev. Lett.}\ }\textbf {\bibinfo {volume} {93}},\
  \bibinfo {pages} {200404} (\bibinfo {year} {2004})}\BibitemShut {NoStop}%
\bibitem [{\citenamefont {Boninsegni}\ \emph {et~al.}(2006)\citenamefont
  {Boninsegni}, \citenamefont {Prokof'ev},\ and\ \citenamefont
  {Svistunov}}]{Boninsegni2006}%
  \BibitemOpen
  \bibfield  {author} {\bibinfo {author} {\bibfnamefont {M.}~\bibnamefont
  {Boninsegni}}, \bibinfo {author} {\bibfnamefont {N.~V.}\ \bibnamefont
  {Prokof'ev}}, \ and\ \bibinfo {author} {\bibfnamefont {B.~V.}\ \bibnamefont
  {Svistunov}},\ }\href {\doibase 10.1103/PhysRevE.74.036701} {\bibfield
  {journal} {\bibinfo  {journal} {Phys. Rev. E}\ }\textbf {\bibinfo {volume}
  {74}},\ \bibinfo {pages} {036701} (\bibinfo {year} {2006})}\BibitemShut
  {NoStop}%
\bibitem [{\citenamefont {Ludwig}\ \emph {et~al.}(2011)\citenamefont {Ludwig},
  \citenamefont {Floerchinger}, \citenamefont {Moroz},\ and\ \citenamefont
  {Wetterich}}]{Ludwig2011}%
  \BibitemOpen
  \bibfield  {author} {\bibinfo {author} {\bibfnamefont {D.}~\bibnamefont
  {Ludwig}}, \bibinfo {author} {\bibfnamefont {S.}~\bibnamefont
  {Floerchinger}}, \bibinfo {author} {\bibfnamefont {S.}~\bibnamefont {Moroz}},
  \ and\ \bibinfo {author} {\bibfnamefont {C.}~\bibnamefont {Wetterich}},\
  }\href {\doibase 10.1103/PhysRevA.84.033629} {\bibfield  {journal} {\bibinfo
  {journal} {Phys. Rev. A}\ }\textbf {\bibinfo {volume} {84}},\ \bibinfo
  {pages} {033629} (\bibinfo {year} {2011})}\BibitemShut {NoStop}%
\bibitem [{\citenamefont {Idziaszek}\ and\ \citenamefont
  {Karwasz}(2009)}]{Idziaszek2009}%
  \BibitemOpen
  \bibfield  {author} {\bibinfo {author} {\bibfnamefont {Z.}~\bibnamefont
  {Idziaszek}}\ and\ \bibinfo {author} {\bibfnamefont {G.}~\bibnamefont
  {Karwasz}},\ }\href {\doibase 10.1140/epjd/e2009-00028-6} {\bibfield
  {journal} {\bibinfo  {journal} {Eur. Phys. J. D}\ }\textbf {\bibinfo {volume}
  {51}},\ \bibinfo {pages} {347} (\bibinfo {year} {2009})}\BibitemShut
  {NoStop}%
\bibitem [{\citenamefont {Greene}\ \emph {et~al.}(2000)\citenamefont {Greene},
  \citenamefont {Dickinson},\ and\ \citenamefont
  {Sadeghpour}}]{greene_creation_2000}%
  \BibitemOpen
  \bibfield  {author} {\bibinfo {author} {\bibfnamefont {C.~H.}\ \bibnamefont
  {Greene}}, \bibinfo {author} {\bibfnamefont {A.~S.}\ \bibnamefont
  {Dickinson}}, \ and\ \bibinfo {author} {\bibfnamefont {H.~R.}\ \bibnamefont
  {Sadeghpour}},\ }\href
  {http://journals.aps.org/prl/abstract/10.1103/PhysRevLett.85.2458} {\bibfield
   {journal} {\bibinfo  {journal} {Phys. Rev. Lett.}\ }\textbf {\bibinfo
  {volume} {85}},\ \bibinfo {pages} {2458} (\bibinfo {year}
  {2000})}\BibitemShut {NoStop}%
\bibitem [{\citenamefont {Bendkowsky}\ \emph {et~al.}(2009)\citenamefont
  {Bendkowsky}, \citenamefont {Butscher}, \citenamefont {Nipper}, \citenamefont
  {Shaffer}, \citenamefont {L{\"o}w},\ and\ \citenamefont
  {Pfau}}]{bendkowsky_observation_2009}%
  \BibitemOpen
  \bibfield  {author} {\bibinfo {author} {\bibfnamefont {V.}~\bibnamefont
  {Bendkowsky}}, \bibinfo {author} {\bibfnamefont {B.}~\bibnamefont
  {Butscher}}, \bibinfo {author} {\bibfnamefont {J.}~\bibnamefont {Nipper}},
  \bibinfo {author} {\bibfnamefont {J.~P.}\ \bibnamefont {Shaffer}}, \bibinfo
  {author} {\bibfnamefont {R.}~\bibnamefont {L{\"o}w}}, \ and\ \bibinfo
  {author} {\bibfnamefont {T.}~\bibnamefont {Pfau}},\ }\href {\doibase
  10.1038/nature07945} {\bibfield  {journal} {\bibinfo  {journal} {Nature}\
  }\textbf {\bibinfo {volume} {458}},\ \bibinfo {pages} {1005} (\bibinfo {year}
  {2009})}\BibitemShut {NoStop}%
\bibitem [{\citenamefont {Pedersen}(2016)}]{Pedersen_Exciton_Stark_shift_2016}%
  \BibitemOpen
  \bibfield  {author} {\bibinfo {author} {\bibfnamefont {T.~G.}\ \bibnamefont
  {Pedersen}},\ }\href {\doibase 10.1103/PhysRevB.94.125424} {\bibfield
  {journal} {\bibinfo  {journal} {Phys. Rev. B}\ }\textbf {\bibinfo {volume}
  {94}},\ \bibinfo {pages} {125424} (\bibinfo {year} {2016})}\BibitemShut
  {NoStop}%
\bibitem [{\citenamefont {Chevy}(2006)}]{Chevy2006}%
  \BibitemOpen
  \bibfield  {author} {\bibinfo {author} {\bibfnamefont {F.}~\bibnamefont
  {Chevy}},\ }\href {\doibase 10.1103/PhysRevA.74.063628} {\bibfield  {journal}
  {\bibinfo  {journal} {Phys. Rev. A}\ }\textbf {\bibinfo {volume} {74}},\
  \bibinfo {pages} {063628} (\bibinfo {year} {2006})}\BibitemShut {NoStop}%
\bibitem [{\citenamefont {Prokof'ev}\ and\ \citenamefont
  {Svistunov}(2008)}]{Prokofev_Svistunov_PRB_2008}%
  \BibitemOpen
  \bibfield  {author} {\bibinfo {author} {\bibfnamefont {N.}~\bibnamefont
  {Prokof'ev}}\ and\ \bibinfo {author} {\bibfnamefont {B.}~\bibnamefont
  {Svistunov}},\ }\href {\doibase 10.1103/PhysRevB.77.020408} {\bibfield
  {journal} {\bibinfo  {journal} {Phys. Rev. B}\ }\textbf {\bibinfo {volume}
  {77}},\ \bibinfo {pages} {020408} (\bibinfo {year} {2008})}\BibitemShut
  {NoStop}%
\bibitem [{\citenamefont {Schmidt}\ and\ \citenamefont
  {Enss}(2011)}]{Schmidt_Enss_PRA_2011}%
  \BibitemOpen
  \bibfield  {author} {\bibinfo {author} {\bibfnamefont {R.}~\bibnamefont
  {Schmidt}}\ and\ \bibinfo {author} {\bibfnamefont {T.}~\bibnamefont {Enss}},\
  }\href {\doibase 10.1103/PhysRevA.83.063620} {\bibfield  {journal} {\bibinfo
  {journal} {Phys. Rev. A}\ }\textbf {\bibinfo {volume} {83}},\ \bibinfo
  {pages} {063620} (\bibinfo {year} {2011})}\BibitemShut {NoStop}%
\bibitem [{\citenamefont {Parish}\ and\ \citenamefont
  {Levinsen}(2013)}]{Parish_Meera_PRA_2013}%
  \BibitemOpen
  \bibfield  {author} {\bibinfo {author} {\bibfnamefont {M.~M.}\ \bibnamefont
  {Parish}}\ and\ \bibinfo {author} {\bibfnamefont {J.}~\bibnamefont
  {Levinsen}},\ }\href {\doibase 10.1103/PhysRevA.87.033616} {\bibfield
  {journal} {\bibinfo  {journal} {Phys. Rev. A}\ }\textbf {\bibinfo {volume}
  {87}},\ \bibinfo {pages} {033616} (\bibinfo {year} {2013})}\BibitemShut
  {NoStop}%
\bibitem [{\citenamefont {Massignan}\ \emph {et~al.}(2014)\citenamefont
  {Massignan}, \citenamefont {Zaccanti},\ and\ \citenamefont
  {Bruun}}]{Massignan2014}%
  \BibitemOpen
  \bibfield  {author} {\bibinfo {author} {\bibfnamefont {P.}~\bibnamefont
  {Massignan}}, \bibinfo {author} {\bibfnamefont {M.}~\bibnamefont {Zaccanti}},
  \ and\ \bibinfo {author} {\bibfnamefont {G.~M.}\ \bibnamefont {Bruun}},\
  }\href {http://stacks.iop.org/0034-4885/77/i=3/a=034401} {\bibfield
  {journal} {\bibinfo  {journal} {Rep. Prog. Phys.}\ }\textbf {\bibinfo
  {volume} {77}},\ \bibinfo {pages} {034401} (\bibinfo {year}
  {2014})}\BibitemShut {NoStop}%
\bibitem [{\citenamefont {Rosch}(1999)}]{Rosch_1999}%
  \BibitemOpen
  \bibfield  {author} {\bibinfo {author} {\bibfnamefont {A.}~\bibnamefont
  {Rosch}},\ }\href {\doibase 10.1080/000187399243446} {\bibfield  {journal}
  {\bibinfo  {journal} {Adv. Phys.}\ }\textbf {\bibinfo {volume} {48}},\
  \bibinfo {pages} {295} (\bibinfo {year} {1999})}\BibitemShut {NoStop}%
\bibitem [{\citenamefont {Koschorreck}\ \emph {et~al.}(2012)\citenamefont
  {Koschorreck}, \citenamefont {Pertot}, \citenamefont {Vogt}, \citenamefont
  {Fr\"{o}hlich}, \citenamefont {Feld},\ and\ \citenamefont
  {K\"{o}hl}}]{Koschorreck2012}%
  \BibitemOpen
  \bibfield  {author} {\bibinfo {author} {\bibfnamefont {M.}~\bibnamefont
  {Koschorreck}}, \bibinfo {author} {\bibfnamefont {D.}~\bibnamefont {Pertot}},
  \bibinfo {author} {\bibfnamefont {E.}~\bibnamefont {Vogt}}, \bibinfo {author}
  {\bibfnamefont {B.}~\bibnamefont {Fr\"{o}hlich}}, \bibinfo {author}
  {\bibfnamefont {M.}~\bibnamefont {Feld}}, \ and\ \bibinfo {author}
  {\bibfnamefont {M.}~\bibnamefont {K\"{o}hl}},\ }\href {\doibase
  10.1038/nature11151} {\bibfield  {journal} {\bibinfo  {journal} {Nature}\
  }\textbf {\bibinfo {volume} {485}},\ \bibinfo {pages} {619} (\bibinfo {year}
  {2012})}\BibitemShut {NoStop}%
\bibitem [{\citenamefont {Fumi}(1955)}]{fumi_1955}%
  \BibitemOpen
  \bibfield  {author} {\bibinfo {author} {\bibfnamefont {F.~G.}\ \bibnamefont
  {Fumi}},\ }\href {\doibase 10.1080/14786440908520622} {\bibfield  {journal}
  {\bibinfo  {journal} {Philos. Mag.}\ }\textbf {\bibinfo {volume} {46}},\
  \bibinfo {pages} {1007} (\bibinfo {year} {1955})}\BibitemShut {NoStop}%
\bibitem [{\citenamefont {Pimenov}\ and\ \citenamefont
  {Goldstein}(2018)}]{Pimenov_2018}%
  \BibitemOpen
  \bibfield  {author} {\bibinfo {author} {\bibfnamefont {D.}~\bibnamefont
  {Pimenov}}\ and\ \bibinfo {author} {\bibfnamefont {M.}~\bibnamefont
  {Goldstein}},\ }\href {\doibase 10.1103/PhysRevB.98.220302} {\bibfield
  {journal} {\bibinfo  {journal} {Phys. Rev. B}\ }\textbf {\bibinfo {volume}
  {98}},\ \bibinfo {pages} {220302} (\bibinfo {year} {2018})}\BibitemShut
  {NoStop}%
\bibitem [{\citenamefont {Tiene}\ \emph {et~al.}(2019)\citenamefont {Tiene},
  \citenamefont {Levinsen}, \citenamefont {Parish}, \citenamefont {MacDonald},
  \citenamefont {Keeling},\ and\ \citenamefont
  {Marchetti}}]{Tiene_Levinsen_Parish_MacDonald_Keeling_Marchetti_2019}%
  \BibitemOpen
  \bibfield  {author} {\bibinfo {author} {\bibfnamefont {A.}~\bibnamefont
  {Tiene}}, \bibinfo {author} {\bibfnamefont {J.}~\bibnamefont {Levinsen}},
  \bibinfo {author} {\bibfnamefont {M.~M.}\ \bibnamefont {Parish}}, \bibinfo
  {author} {\bibfnamefont {A.~H.}\ \bibnamefont {MacDonald}}, \bibinfo {author}
  {\bibfnamefont {J.}~\bibnamefont {Keeling}}, \ and\ \bibinfo {author}
  {\bibfnamefont {F.~M.}\ \bibnamefont {Marchetti}},\ }\href
  {http://arxiv.org/abs/1911.08808} {\bibfield  {journal} {\bibinfo  {journal}
  {arXiv: 1911.08808}\ } (\bibinfo {year} {2019})}\BibitemShut {NoStop}%
\bibitem [{\citenamefont {Semina}\ \emph {et~al.}(2008)\citenamefont {Semina},
  \citenamefont {Sergeev},\ and\ \citenamefont {Suris}}]{semina2008}%
  \BibitemOpen
  \bibfield  {author} {\bibinfo {author} {\bibfnamefont {M.}~\bibnamefont
  {Semina}}, \bibinfo {author} {\bibfnamefont {R.}~\bibnamefont {Sergeev}}, \
  and\ \bibinfo {author} {\bibfnamefont {R.}~\bibnamefont {Suris}},\ }\href
  {\doibase https://doi.org/10.1016/j.physe.2007.09.008} {\bibfield  {journal}
  {\bibinfo  {journal} {Physica E}\ }\textbf {\bibinfo {volume} {40}},\
  \bibinfo {pages} {1357 } (\bibinfo {year} {2008})}\BibitemShut {NoStop}%
\bibitem [{\citenamefont {Filikhin}\ \emph {et~al.}(2018)\citenamefont
  {Filikhin}, \citenamefont {Kezerashvili}, \citenamefont {Tsiklauri},\ and\
  \citenamefont {Vlahovic}}]{filikhin2018trions}%
  \BibitemOpen
  \bibfield  {author} {\bibinfo {author} {\bibfnamefont {I.}~\bibnamefont
  {Filikhin}}, \bibinfo {author} {\bibfnamefont {R.~Y.}\ \bibnamefont
  {Kezerashvili}}, \bibinfo {author} {\bibfnamefont {S.~M.}\ \bibnamefont
  {Tsiklauri}}, \ and\ \bibinfo {author} {\bibfnamefont {B.}~\bibnamefont
  {Vlahovic}},\ }\href {\doibase 10.1088/1361-6528/aaa94d} {\bibfield
  {journal} {\bibinfo  {journal} {Nanotechnology}\ }\textbf {\bibinfo {volume}
  {29}},\ \bibinfo {pages} {124002} (\bibinfo {year} {2018})}\BibitemShut
  {NoStop}%
\bibitem [{\citenamefont {Combescot}\ and\ \citenamefont
  {Betbeder-Matibet}(2011)}]{Combescot2011}%
  \BibitemOpen
  \bibfield  {author} {\bibinfo {author} {\bibfnamefont {M.}~\bibnamefont
  {Combescot}}\ and\ \bibinfo {author} {\bibfnamefont {O.}~\bibnamefont
  {Betbeder-Matibet}},\ }\href {\doibase 10.1140/epjb/e2010-10804-6} {\bibfield
   {journal} {\bibinfo  {journal} {Eur. Phys. J. B}\ }\textbf {\bibinfo
  {volume} {79}},\ \bibinfo {pages} {401} (\bibinfo {year} {2011})}\BibitemShut
  {NoStop}%
\bibitem [{\citenamefont {Stier}\ \emph {et~al.}(2018)\citenamefont {Stier},
  \citenamefont {Wilson}, \citenamefont {Velizhanin}, \citenamefont {Kono},
  \citenamefont {Xu},\ and\ \citenamefont {Crooker}}]{Stier_2018}%
  \BibitemOpen
  \bibfield  {author} {\bibinfo {author} {\bibfnamefont {A.}~\bibnamefont
  {Stier}}, \bibinfo {author} {\bibfnamefont {N.}~\bibnamefont {Wilson}},
  \bibinfo {author} {\bibfnamefont {K.}~\bibnamefont {Velizhanin}}, \bibinfo
  {author} {\bibfnamefont {J.}~\bibnamefont {Kono}}, \bibinfo {author}
  {\bibfnamefont {X.}~\bibnamefont {Xu}}, \ and\ \bibinfo {author}
  {\bibfnamefont {S.}~\bibnamefont {Crooker}},\ }\href {\doibase
  10.1103/PhysRevLett.120.057405} {\bibfield  {journal} {\bibinfo  {journal}
  {Phys. Rev. Lett.}\ }\textbf {\bibinfo {volume} {120}},\ \bibinfo {pages}
  {057405} (\bibinfo {year} {2018})}\BibitemShut {NoStop}%
\bibitem [{\citenamefont {Goryca}\ \emph {et~al.}(2019)\citenamefont {Goryca},
  \citenamefont {Li}, \citenamefont {Stier}, \citenamefont {Taniguchi},
  \citenamefont {Watanabe}, \citenamefont {Courtade}, \citenamefont {Shree},
  \citenamefont {Robert}, \citenamefont {Urbaszek}, \citenamefont {Marie},\
  and\ \citenamefont {et~al.}}]{Goryca_Stier_2019}%
  \BibitemOpen
  \bibfield  {author} {\bibinfo {author} {\bibfnamefont {M.}~\bibnamefont
  {Goryca}}, \bibinfo {author} {\bibfnamefont {J.}~\bibnamefont {Li}}, \bibinfo
  {author} {\bibfnamefont {A.~V.}\ \bibnamefont {Stier}}, \bibinfo {author}
  {\bibfnamefont {T.}~\bibnamefont {Taniguchi}}, \bibinfo {author}
  {\bibfnamefont {K.}~\bibnamefont {Watanabe}}, \bibinfo {author}
  {\bibfnamefont {E.}~\bibnamefont {Courtade}}, \bibinfo {author}
  {\bibfnamefont {S.}~\bibnamefont {Shree}}, \bibinfo {author} {\bibfnamefont
  {C.}~\bibnamefont {Robert}}, \bibinfo {author} {\bibfnamefont
  {B.}~\bibnamefont {Urbaszek}}, \bibinfo {author} {\bibfnamefont
  {X.}~\bibnamefont {Marie}}, \ and\ \bibinfo {author} {\bibnamefont
  {et~al.}},\ }\href {\doibase 10.1038/s41467-019-12180-y} {\bibfield
  {journal} {\bibinfo  {journal} {Nature Commun.}\ }\textbf {\bibinfo {volume}
  {10}},\ \bibinfo {pages} {4172} (\bibinfo {year} {2019})}\BibitemShut
  {NoStop}%
\bibitem [{\citenamefont {Klein}\ \emph {et~al.}(2019)\citenamefont {Klein},
  \citenamefont {Lorke}, \citenamefont {Florian}, \citenamefont {Sigger},
  \citenamefont {Sigl}, \citenamefont {Rey}, \citenamefont {Wierzbowski},
  \citenamefont {Cerne}, \citenamefont {Müller}, \citenamefont
  {Mitterreiter},\ and\ \citenamefont {et~al.}}]{Klein_2019}%
  \BibitemOpen
  \bibfield  {author} {\bibinfo {author} {\bibfnamefont {J.}~\bibnamefont
  {Klein}}, \bibinfo {author} {\bibfnamefont {M.}~\bibnamefont {Lorke}},
  \bibinfo {author} {\bibfnamefont {M.}~\bibnamefont {Florian}}, \bibinfo
  {author} {\bibfnamefont {F.}~\bibnamefont {Sigger}}, \bibinfo {author}
  {\bibfnamefont {L.}~\bibnamefont {Sigl}}, \bibinfo {author} {\bibfnamefont
  {S.}~\bibnamefont {Rey}}, \bibinfo {author} {\bibfnamefont {J.}~\bibnamefont
  {Wierzbowski}}, \bibinfo {author} {\bibfnamefont {J.}~\bibnamefont {Cerne}},
  \bibinfo {author} {\bibfnamefont {K.}~\bibnamefont {Müller}}, \bibinfo
  {author} {\bibfnamefont {E.}~\bibnamefont {Mitterreiter}}, \ and\ \bibinfo
  {author} {\bibnamefont {et~al.}},\ }\href {\doibase
  10.1038/s41467-019-10632-z} {\bibfield  {journal} {\bibinfo  {journal}
  {Nature Commun.}\ }\textbf {\bibinfo {volume} {10}},\ \bibinfo {pages} {2755}
  (\bibinfo {year} {2019})}\BibitemShut {NoStop}%
\bibitem [{\citenamefont {Petrov}\ \emph {et~al.}(2004)\citenamefont {Petrov},
  \citenamefont {Salomon},\ and\ \citenamefont {Shlyapnikov}}]{Petrov2004}%
  \BibitemOpen
  \bibfield  {author} {\bibinfo {author} {\bibfnamefont {D.~S.}\ \bibnamefont
  {Petrov}}, \bibinfo {author} {\bibfnamefont {C.}~\bibnamefont {Salomon}}, \
  and\ \bibinfo {author} {\bibfnamefont {G.~V.}\ \bibnamefont {Shlyapnikov}},\
  }\href {\doibase 10.1103/PhysRevLett.93.090404} {\bibfield  {journal}
  {\bibinfo  {journal} {Phys. Rev. Lett.}\ }\textbf {\bibinfo {volume} {93}},\
  \bibinfo {pages} {090404} (\bibinfo {year} {2004})}\BibitemShut {NoStop}%
\bibitem [{\citenamefont {Fey}\ \emph {et~al.}(2019)\citenamefont {Fey},
  \citenamefont {Yang}, \citenamefont {Rittenhouse}, \citenamefont {Munkes},
  \citenamefont {Baluktsian}, \citenamefont {Schmelcher}, \citenamefont
  {Sadeghpour},\ and\ \citenamefont {Shaffer}}]{Fey2019}%
  \BibitemOpen
  \bibfield  {author} {\bibinfo {author} {\bibfnamefont {C.}~\bibnamefont
  {Fey}}, \bibinfo {author} {\bibfnamefont {J.}~\bibnamefont {Yang}}, \bibinfo
  {author} {\bibfnamefont {S.~T.}\ \bibnamefont {Rittenhouse}}, \bibinfo
  {author} {\bibfnamefont {F.}~\bibnamefont {Munkes}}, \bibinfo {author}
  {\bibfnamefont {M.}~\bibnamefont {Baluktsian}}, \bibinfo {author}
  {\bibfnamefont {P.}~\bibnamefont {Schmelcher}}, \bibinfo {author}
  {\bibfnamefont {H.~R.}\ \bibnamefont {Sadeghpour}}, \ and\ \bibinfo {author}
  {\bibfnamefont {J.~P.}\ \bibnamefont {Shaffer}},\ }\href {\doibase
  10.1103/PhysRevLett.122.103001} {\bibfield  {journal} {\bibinfo  {journal}
  {Phys. Rev. Lett.}\ }\textbf {\bibinfo {volume} {122}},\ \bibinfo {pages}
  {103001} (\bibinfo {year} {2019})}\BibitemShut {NoStop}%
\bibitem [{\citenamefont {Mirgorodskiy}\ \emph {et~al.}(2017)\citenamefont
  {Mirgorodskiy}, \citenamefont {Christaller}, \citenamefont {Braun},
  \citenamefont {Paris-Mandoki}, \citenamefont {Tresp},\ and\ \citenamefont
  {Hofferberth}}]{Mirgorodskiy2017}%
  \BibitemOpen
  \bibfield  {author} {\bibinfo {author} {\bibfnamefont {I.}~\bibnamefont
  {Mirgorodskiy}}, \bibinfo {author} {\bibfnamefont {F.}~\bibnamefont
  {Christaller}}, \bibinfo {author} {\bibfnamefont {C.}~\bibnamefont {Braun}},
  \bibinfo {author} {\bibfnamefont {A.}~\bibnamefont {Paris-Mandoki}}, \bibinfo
  {author} {\bibfnamefont {C.}~\bibnamefont {Tresp}}, \ and\ \bibinfo {author}
  {\bibfnamefont {S.}~\bibnamefont {Hofferberth}},\ }\href {\doibase
  10.1103/PhysRevA.96.011402} {\bibfield  {journal} {\bibinfo  {journal} {Phys.
  Rev. A}\ }\textbf {\bibinfo {volume} {96}},\ \bibinfo {pages} {011402}
  (\bibinfo {year} {2017})}\BibitemShut {NoStop}%
\bibitem [{\citenamefont {Levitov}\ \emph {et~al.}(1996)\citenamefont
  {Levitov}, \citenamefont {Lee},\ and\ \citenamefont {Lesovik}}]{Levitov1996}%
  \BibitemOpen
  \bibfield  {author} {\bibinfo {author} {\bibfnamefont {L.~S.}\ \bibnamefont
  {Levitov}}, \bibinfo {author} {\bibfnamefont {H.}~\bibnamefont {Lee}}, \ and\
  \bibinfo {author} {\bibfnamefont {G.~B.}\ \bibnamefont {Lesovik}},\ }\href
  {\doibase 10.1063/1.531672} {\bibfield  {journal} {\bibinfo  {journal} {J.
  Math. Phys.}\ }\textbf {\bibinfo {volume} {37}},\ \bibinfo {pages} {4845}
  (\bibinfo {year} {1996})}\BibitemShut {NoStop}%
\bibitem [{\citenamefont {Klich}(2002)}]{Klich_2002}%
  \BibitemOpen
  \bibfield  {author} {\bibinfo {author} {\bibfnamefont {I.}~\bibnamefont
  {Klich}},\ }\href {http://arxiv.org/abs/cond-mat/0209642} {\bibfield
  {journal} {\bibinfo  {journal} {arXiv cond-mat/0209642}\ } (\bibinfo {year}
  {2002})}\BibitemShut {NoStop}%
\bibitem [{\citenamefont {Sch\"onhammer}(2007)}]{Schoenhammer_PRB_2007}%
  \BibitemOpen
  \bibfield  {author} {\bibinfo {author} {\bibfnamefont {K.}~\bibnamefont
  {Sch\"onhammer}},\ }\href {\doibase 10.1103/PhysRevB.75.205329} {\bibfield
  {journal} {\bibinfo  {journal} {Phys. Rev. B}\ }\textbf {\bibinfo {volume}
  {75}},\ \bibinfo {pages} {205329} (\bibinfo {year} {2007})}\BibitemShut
  {NoStop}%
\bibitem [{\citenamefont {Kain}\ and\ \citenamefont
  {Ling}(2017)}]{Kain_Ling_2017}%
  \BibitemOpen
  \bibfield  {author} {\bibinfo {author} {\bibfnamefont {B.}~\bibnamefont
  {Kain}}\ and\ \bibinfo {author} {\bibfnamefont {H.~Y.}\ \bibnamefont
  {Ling}},\ }\href {\doibase 10.1103/PhysRevA.96.033627} {\bibfield  {journal}
  {\bibinfo  {journal} {Phys. Rev. A}\ }\textbf {\bibinfo {volume} {96}},\
  \bibinfo {pages} {033627} (\bibinfo {year} {2017})}\BibitemShut {NoStop}%
\bibitem [{\citenamefont {Cetina}\ \emph {et~al.}(2016)\citenamefont {Cetina},
  \citenamefont {Jag}, \citenamefont {Lous}, \citenamefont {Fritsche},
  \citenamefont {Walraven}, \citenamefont {Grimm}, \citenamefont {Levinsen},
  \citenamefont {Parish}, \citenamefont {Schmidt}, \citenamefont {Knap},\ and\
  \citenamefont {Demler}}]{cetina_ultrafast_2016}%
  \BibitemOpen
  \bibfield  {author} {\bibinfo {author} {\bibfnamefont {M.}~\bibnamefont
  {Cetina}}, \bibinfo {author} {\bibfnamefont {M.}~\bibnamefont {Jag}},
  \bibinfo {author} {\bibfnamefont {R.~S.}\ \bibnamefont {Lous}}, \bibinfo
  {author} {\bibfnamefont {I.}~\bibnamefont {Fritsche}}, \bibinfo {author}
  {\bibfnamefont {J.~T.~M.}\ \bibnamefont {Walraven}}, \bibinfo {author}
  {\bibfnamefont {R.}~\bibnamefont {Grimm}}, \bibinfo {author} {\bibfnamefont
  {J.}~\bibnamefont {Levinsen}}, \bibinfo {author} {\bibfnamefont {M.~M.}\
  \bibnamefont {Parish}}, \bibinfo {author} {\bibfnamefont {R.}~\bibnamefont
  {Schmidt}}, \bibinfo {author} {\bibfnamefont {M.}~\bibnamefont {Knap}}, \
  and\ \bibinfo {author} {\bibfnamefont {E.}~\bibnamefont {Demler}},\ }\href
  {\doibase 10.1126/science.aaf5134} {\bibfield  {journal} {\bibinfo  {journal}
  {Science}\ }\textbf {\bibinfo {volume} {354}},\ \bibinfo {pages} {96}
  (\bibinfo {year} {2016})}\BibitemShut {NoStop}%
\end{thebibliography}
%

\onecolumngrid
\appendix

\section{Kinetic Hamiltonian in valence coordinates} 
\label{sec:appendix}
We consider three particles ($i=1,2,3$) in two dimensions with masses $m_i$ and coordinates $\vec{R}_i$. The kinetic Hamiltonian reads 
 
\begin{equation}
\hat{H}_\text{kin} = - \sum_{i=1}^3 \frac{\hbar^2}{2m_i} \Delta_{\vec{R}_i} .
\label{eqn:Hkin_cartesian}
\end{equation}

In order to separate the center of mass motion we transform $\vec{r}_1=\vec{R}_1-\vec{R}_3$, $\vec{r}_2=\vec{R}_2-\vec{R}_3$ and $\vec{r}_\text{COM}=\frac{1}{M}(m_1 \vec{R}_1 + m_2 \vec{R}_2 +m_3 \vec{R}_3)$. The resulting kinetic Hamiltonian reads
\begin{equation}
\begin{split}
\hat{H}_\text{kin} &= -\frac{\hbar^2}{2\mu_1} \Delta_{\vec{r}_1}  -\frac{\hbar^2}{2\mu_2} \Delta_{\vec{r}_2} -\frac{\hbar^2}{2 m_3} \vec{\nabla}_{\vec{r}_1} \cdot \vec{\nabla}_{\vec{r}_2} -\frac{\hbar^2}{2 M}  \Delta_{\vec{r}_\text{COM}}
 \end{split}
\label{eqn:Hkin_valence}
\end{equation}
with the total mass $M=m_1+m_2+m_3$ and the reduced masses $\mu_1=m_1 m_3/(m_1+m_3)$ and $\mu_2=m_2 m_3/(m_2+m_3)$.
The center of mass motion can be separated since all interaction potentials are independent of $\vec{r}_\text{COM}$.

Next we parametrize the internal degrees of freedom  
$\vec{r_1}=r_1 [\cos(\alpha+ \frac{\theta}{2}) \vec{e}_x +\sin(\alpha+\frac{\theta}{2}) \vec{e}_y]$ and $\vec{r_2}=r_2 [\cos(\alpha- \frac{\theta}{2}) \vec{e}_x +\sin(\alpha-\frac{\theta}{2}) \vec{e}_y]$ via the two radii $r_1$, $r_2$ and the two angles $\theta$ and $\alpha$, see Fig \ref{fig:coordinates} (a). The resulting kinetic Hamiltonian (without center of mass motion) is
\begin{align}
\hat{H}^\text{rovib}_\text{kin}  &=-\frac{\hbar^2}{2 \mu_1} \left(\partial^2_{r_1} + \frac{\partial_{r_1}}{r_1} +\frac{ \partial^2_\theta}{r_1^2}\right) -\frac{\hbar^2}{2 \mu_2} \left(\partial^2_{r_2} + \frac{\partial_{r_2}}{r_2} +\frac{ \partial^2_\theta}{r_2^2}\right)
 \nonumber \\
&- \frac{\hbar^2}{m_3} \left[ \cos \theta \partial_{r_1} \partial_{r_2} - \frac{\cos \theta \partial_\theta^2}{r_1 r_2}  - \left(\frac{\partial_{r_1}}{r_2} + \frac{\partial_{r_2}}{r_1} \right) \sin \theta \partial_\theta \right] \nonumber\\
&- \frac{\hbar^2}{2 \mu_1}\left(\frac{\partial^2_\alpha}{4 r_1^2} - \frac{\partial_\theta \partial_\alpha}{r_1^2} \right)
- \frac{\hbar^2}{2 \mu_2}\left(\frac{\partial^2_\alpha}{4 r_2^2}  + \frac{\partial_\theta \partial_\alpha}{r_2^2}  \right) \nonumber \\
&-\frac{\hbar^2}{2m_3} \left[ \frac{\cos \theta}{2 r_1 r_2} \partial_\alpha^2 +\sin \theta \left( \frac{\partial_{r_2} \partial_\alpha}{r_1}-\frac{\partial_{r_1} \partial_\alpha}{r_2} \right)  \right]
\label{eqn:Hkin_rovib_valence}
\end{align} 
In the case of $ m_1=m_2$ the kinetic Hamiltonian is invariant under exchange of $\vec{r}_1$ and $\vec{r}_2$ which corresponds to the transformation $r_1 \mapsto r_2$, $r_2 \mapsto r_1$ and  $\theta \mapsto -\theta$.

If all interactions do not depend on the angle $\alpha$, it is convenient to express the wave function as $\psi(r_1,r_2,\theta,\alpha)= \frac{u(r_1,r_2,\theta)}{\sqrt{2 \pi r_1 r_2}} \exp(i m \alpha)$ where $m$ is a conserved angular momentum quantum number. The normalization condition is
\begin{equation}
\int \limits_{r_1=0}^\infty \int \limits_{r_2=0}^\infty \int \limits_{\theta=0}^{2 \pi} \int \limits_{\alpha=0}^{2 \pi} dr_1 dr_2 \ d\theta d \alpha r_1 r_2 |\psi(r_1,r_2,\theta,\alpha)|^2 =\int \limits_{r_1=0}^\infty \int \limits_{r_2=0}^\infty \int \limits_{\theta=0}^{2 \pi}  dr_1 dr_2 \ d\theta |u(r_1,r_2,\theta)|^2 = 1    
\end{equation}
The purely vibrational ($m=0$) Hamiltonian which acts on $u(r_1,r_2,\theta)$ is, finally, given by

\begin{equation}
\begin{split}
\hat{H}^\text{vib}_\text{kin} =&-\frac{\hbar^2}{2 \mu_1} \left(\partial^2_{r_1} + \frac{1}{4 r_1^2} + \frac{\partial^2_{\theta}}{r_1^2} \right) -\frac{\hbar^2}{2 \mu_2} \left(\partial^2_{r_2} + \frac{1}{4 r_2^2} + \frac{\partial^2_{\theta}}{r_2^2} \right)  \nonumber \\
&- \frac{\hbar^2}{m_3} \left[
\frac{1}{r_1 r_2} \left(\frac{1}{4} \cos \theta - \partial_\theta \cos \theta \partial_\theta \right)
+\cos \theta \partial_{r_1} \partial_{r_2}  - \frac{1}{2} \left(\frac{\partial_{r_1}}{r_2} + \frac{\partial_{r_2}}{r_1} \right)  \left(\sin \theta \partial_\theta +  \partial_\theta \sin \theta \right) \right].
\end{split}
\label{eqn:Hkin_vib_valence}
\end{equation}

\section{Polaron absorption spectrum from a functional determinant approach}
\label{sec:appendix_FDA}
To predict the spectral function $A(E)$ shown in Fig.~\ref{fig:sidler}, we employ a functional determinant approach (FDA) which is exact for bilinear Hamiltonians and relies on a mapping of expectation values of exponentiated many-body operators to determinants of single-particle operators \cite{Levitov1996, Klich_2002, Schoenhammer_PRB_2007, schmidt2017}.
We employ the Hamiltonian
\begin{equation}
\hat H=\sum_\veck \frac{k^2}{2\mu_{Xe}}  \ced_\veck \ce_\veck  + \frac{1}{\mathcal{A}} \sum_{\veck\vecq}V_{Xe}(\vecq)\ced_{\veck+\vecq} \ce_\veck , 
\label{eqn: Hamiltonian_polaron_appendix}
\end{equation}
where $\mathcal{A}$ is the system area, $\mu_{Xe}=(m_h+ 2 m_e)/(m^2_e+m_h m_e)$ is the reduced exciton-electron mass and $V_{Xe}(\vecq)$ is the two-dimensional Fourier transform of the exciton-electron interaction $V^{(S)}_{Xe}(r)$ given in (\ref{eqn:pseudopotential}) with material parameters for MoSe$_2$ as specified in Table \ref{tab:table2}. The Hamiltonian (\ref{eqn: Hamiltonian_polaron_appendix}) is an approximation of the full many-body Hamiltonian, since it neglects non-bilinear couplings terms due to finite mass ratio of the electron and exciton masses \cite{Kain_Ling_2017}. However, $\hat{H}$ becomes exact in the limit $m_h \gg m_e$ and has been shown to be in excellent agreement with polaron spectra of heavy impurities in ultracold fermionic atomic gases \cite{cetina_ultrafast_2016} as well as Rydberg impurities in Bose gases \cite{Camargo2018}.  

Following the FDA, the polaron spectral function $A(E)$ is given by the Fourier transform $A(E,\epsilon_F)=2 \Re(\int^{\infty}_0 S(t) \exp{(iEt)} dt)$ of the Loschmidt echo $S(t)=\det[1-\hat{n}+\hat{n} \exp(i \hat{h}_0t) \exp(-i \hat{h}t)]$ which depends on single-particle operators $\hat{h}_0$, $\hat{h}$ and $\hat{n}$ \cite{schmidt2017}, given by $\hat{h}_0= \vec{k}^2/(2 \mu_{Xe})$ and $\hat{h}=\hat{h}_0 + V_{Xe}^{(S)}(r)$ describing the relative dynamics of a single exciton-electron pair without or with exciton-electron interactions, respectively. Furthermore, $\hat{n}=\theta(\epsilon_F-\hat{h}_0)$ is the occupation operator of the non-interacting Fermi gas at zero temperature, where $\theta$ is the Heaviside step function. 
The time evolution and evaluation of $S(t)$ is performed by diagonalizing $\hat{h}$ and $\hat{h}_0$ in a circular box of radius $10^5 a_0$ in the subspace of conserved zero angular momentum using a radial DVR grid. Finally, the resulting polaron spectrum is shifted according to $A(E) \mapsto A(E-0.8 \epsilon_F+ 1.2 \text{meV}) $ in order to account for the difference between the experimentally observed trion energy of 26.5 meV and our ab initio prediction of 27.7 meV, see Tab.~\ref{tab:table1}, and, in order to include the experimentally determined corrections due to phase space filling, screening and band gap renormalization \cite{Sidler2016}.

\end{document}